%% file: main.tex
\pdfoutput=1
\documentclass[sigconf,natbib=false]{acmart}

\AtBeginDocument{%
  }

\usepackage{threeparttable}
\usepackage{diagbox}
\usepackage{amsmath,amsfonts}

\usepackage{amssymb}
\usepackage{graphicx}
\usepackage{textcomp}
\usepackage{xcolor}
\usepackage{url}
\usepackage{listings}
\usepackage{float} 
\usepackage{multirow}
\usepackage{subfigure}
\usepackage{bm}
\usepackage{fancyhdr}
\usepackage{tcolorbox}
\usepackage{booktabs}
\usepackage{colortbl}
\usepackage{bbding}
\usepackage{pifont}
\usepackage{dutchcal}
\usepackage{ragged2e}
\usepackage{balance}

\DeclareUnicodeCharacter{0301}{*************************************}

\newcommand{\logtext}[1]{\textls[-40]{\texttt{``#1"}}}

\newcommand\tool{LogShrink }
 
\setcopyright{acmcopyright}

\copyrightyear{2024} 
\acmYear{2024} 
\setcopyright{acmlicensed}\acmConference[ICSE 2024]{2024 IEEE/ACM 46th International Conference on Software Engineering}{April 14--20, 2024}{Lisbon, Portugal}
\acmBooktitle{2024 IEEE/ACM 46th International Conference on Software Engineering (ICSE 2024), April 14--20, 2024, Lisbon, Portugal}
\acmPrice{15.00}
\acmDOI{10.1145/3597503.3608129}
\acmISBN{979-8-4007-0217-4/24/04}

\RequirePackage[
  datamodel=acmdatamodel,
  style=acmnumeric,
  sorting=none,
  natbib=false
  ]{biblatex}

\newcommand{\figin}{\vspace{-0.in}}
\newcommand{\tabin}{\vspace{-0.in}}
\begin{document}

\title{LogShrink: Effective Log Compression by Leveraging Commonality and Variability of Log Data}

\author{Xiaoyun Li}
\affiliation{%
  \institution{Sun Yat-sen University}
  \city{Guangzhou}
  \country{China}
}
\email{lixy223@mail2.sysu.edu.cn}

\author{Hongyu Zhang}
\authornote{Hongyu Zhang is the corresponding author.}
\affiliation{%
  \institution{Chongqing University}
  \city{Chongqing}
  \country{China}
}
\email{hyzhang@cqu.edu.cn}

\author{Van-Hoang Le}
\affiliation{%
  \institution{The University of Newcastle}
  \state{NSW}
  \country{Australia}
}
\email{vanhoang.le@uon.edu.au}

\author{Pengfei Chen}
\affiliation{%
  \institution{Sun Yat-sen University}
  \city{Guangzhou}
  \country{China}
}
\email{chenpf7@mail.sysu.edu.cn}


\begin{abstract}
Log data is a crucial resource for recording system events and states during system execution. However, as systems grow in scale, log data generation has become increasingly explosive, leading to an expensive overhead on log storage, such as several petabytes per day in production. To address this issue, log compression has become a crucial task in reducing disk storage while allowing for further log analysis. Unfortunately, existing general-purpose and log-specific compression methods have been limited in their ability to utilize log data characteristics. To overcome these limitations, we conduct an empirical study and obtain three major observations on the characteristics of log data that can facilitate the log compression task. Based on these observations, we propose LogShrink, a novel and effective log compression method by leveraging commonality and variability of log data. An analyzer based on longest common subsequence and entropy techniques is proposed to identify the latent commonality and variability in log messages. The key idea behind this is that the commonality and variability can be exploited to shrink log data with a shorter representation. Besides, a clustering-based sequence sampler is introduced to accelerate the commonality and variability analyzer. The extensive experimental results demonstrate that LogShrink can exceed baselines in compression ratio by 16\% to 356\% on average while preserving a reasonable compression speed. 

\end{abstract}



\maketitle
\vspace{-3pt}
\input{sections/introduction}
\input{sections/motivation}

\input{sections/empirical-study}
\input{sections/methodology}
\input{sections/evaluation}

\input{sections/discussion}

\section{Conclusion}
\label{sec:conclusion}

The sheer volume of log data presents a significant challenge for storage costs. Current compression methods, including general-purpose and log-specific methods, have limited capability in utilizing the characteristics of log data. We have conducted an empirical study on the characteristics of log data and derived three major observations, which led us to propose LogShrink, an effective log compression method, by leveraging commonality and variability of log data. 
Our experimental results show that \tool can outperform existing compressors by 16\% to 356\% on average with respect to compression ratio while maintaining reasonable compression speed. In the future, we will optimize the LogShrink implementation in 
faster programming languages such as C++.


\begin{acks}
We thank all the anonymous reviewers for their insightful reviews. This research is supported by the National Key Research and Development Program of China (2019YFB1804002), the National Natural Science Foundation of China (No.62272495), the Guangdong Basic and Applied Basic Research Foundation (No.2023B1515020054), Australian Research Council Discovery Projects (DP200102940, DP220103044). Xiaoyun Li is supported by the China Scholarship Council (202206380116) during this work.
\end{acks}
\balance
\printbibliography

\end{document}

%% file: sections/introduction.tex
\section{Introduction}
\label{sec:introduction}

Recording system events and status during runtime execution is important for maintaining and troubleshooting software systems. The usage of log data spans a wide range of scopes, including software testing prior to system deployment~\cite{andrews1998testing, chen2018automated}, real-time system performance monitoring~\cite{agrawal2018log, yao2018log4perf}, and root-cause analysis~\cite{he2018identifying, du2017deeplog, ijcai2019-658, zhang2022deeptralog,zhou2019latent}. 
As the scale of a system expands, the volume of log data can increase exponentially. 
Recent reports show that the volume of log data generated from software systems has grown significantly in recent years. A modern software system can produce 
log data at rates of 100 TB to several PB per day~\cite{lin2015cowic, wei2021feasibility}.
In many scenarios like forensic analysis, log data is stored for a long period for backtracking and understanding security issues. For example, system logs must be stored for up to 180 days in AliCloud~\cite{wei2021feasibility}. 
The cost of storing such a vast volume of log data can be prohibitively expensive.
To illustrate, consider a cloud provider that needs to store 1 PB of log data per day, and the cost of logging storage is \$0.50/GiB~\cite{loggingpricing} per month. The monthly bill can reach up to \$465.7k, posing a significant financial burden on storage costs for cloud providers.


To address the challenge of log storage, there are two possible solutions: reducing log generation and compressing log files. Some studies~\cite{ding2015log2,yu2023logreducer,zhao2017log20} propose to generate log messages on demand to reduce the volume of console logs. However, the amount of logs after the reduction process is still significant and often reaches petabyte-level outputs per day (e.g., reducing from 19.7 PB to 12.0 PB per day~\cite{yu2023logreducer}). 
Therefore, log compression is essential to archive large volumes of log data and saves disk storage space while preserving the opportunities for further analysis. Achieving a high compression ratio is critical in practice, as it can result in significant savings in disk space costs.

The most straightforward way to perform log compression is to apply general-purpose data compression algorithms such as lzma~\cite{7za}, gzip~\cite{gzip} and bzip2~\cite{bzip2}. They can obtain a relatively small log file compared to the original log file. These algorithms analyze log files character-by-character, identify the repeated characters in log data, and replace them with a shorter representation. 
However, they cannot fully disclose the redundancy of log files, which are well formatted and might enable more effective compression~\cite{ding2021elise, liu2019logzip}.
To overcome this limitation, some log-specific compression methods~\cite{lin2015cowic, christensen2013adaptive, yao2021improving} are proposed to utilize the latent structure of log data to improve compression results. For example, LogZip~\cite{liu2019logzip} extracts log events and corresponding parameters via iterative clustering and compresses log data using dictionary encoding. LogReducer~\cite{wei2021feasibility} proposes to improve compression ratio through delta encoding of timestamps and elastic encoding of numerical values.
Although effective, these methods are still limited in utilizing the characteristics of log data and their compression ratios can be further improved.

In order to compress log data by leveraging its inherent characteristics, we conduct an empirical study on three real-world log datasets from diverse software systems. 
We obtain three key observations that could potentially enhance both the compression ratio and speed. First, we observe that commonality and variability exist among log messages and they can be exploited to generate a shorter representation for log messages.
Second, we find that the storage style has a significant impact on the compression ratio. Specifically, column-oriented storage style can reduce the compressed file size by 36\% to 103\% compared to traditional row-oriented storage style due to the columnar nature of log data. Lastly, we observe that the distribution of log sequence types is highly imbalanced. Over 50\% of all log sequences belong to only 3\%$\sim$5\% of log sequence types. 



Based on the aforementioned three observations, we propose \textbf{LogShrink}, a novel and effective log-specific compression method by leveraging commonality and variability of log data.
LogShrink first reads and segments the input log file into multiple log chunks to enable parallel batch processing. Each log chunk is initially parsed into structured logs including log headers, log events, and log variables. Next, we propose a novel clustering-based sequence sampling method to extract representative log sequences in each log chunk. Then, we devise an analyzer to identify commonality and variability in log data. The identified characteristics are sent to a compressor, which matches all log data corresponding to their types, encodes values according to their value types, and stores all the content in a column-oriented format. Finally, all the encoded files are compressed using a general-purpose compressor.

We conduct extensive experiments to evaluate LogShrink on 16 public datasets collected from a variety of software systems~\cite{loghubdatasets}. The experimental results show that LogShrink outperforms two log-specific compression methods and three general-purpose compression methods by 16\% to 356\% on average in terms of compression ratios with a reasonable compression speed.
The ablation study on LogShrink confirms that the proposed commonality and variability analyzer and clustering-based sequence sampling contribute to the improvement of both compression ratio and speed. 

In summary, the major contributions of this paper are as follows:

\vspace{-2pt}
\begin{itemize}
    \item We conduct an empirical study on three real-world log datasets and obtain three observations, which can facilitate the log compression task. 
    \item Based on the obtained observations, we propose a novel and effective log compression method by leveraging commonality and variability of log data. Our proposed commonality and variability analyzer can largely improve the compression ratio. In the meantime, a clustering-based sequence sampler can accelerate the analyzing process thus improving the compression speed. 
    \item We perform extensive experiments on 16 public log datasets, which confirm the efficacy of our proposed method. The source code of our tool and our experimental data are available at \url{https://github.com/IntelligentDDS/LogShrink}
\end{itemize}


%% file: sections/motivation.tex
\section{Background and Related Work}
\label{sec:motivation}


\subsection{Background}

\begin{figure}[tbp]
    \centerline{\includegraphics[width=\linewidth]{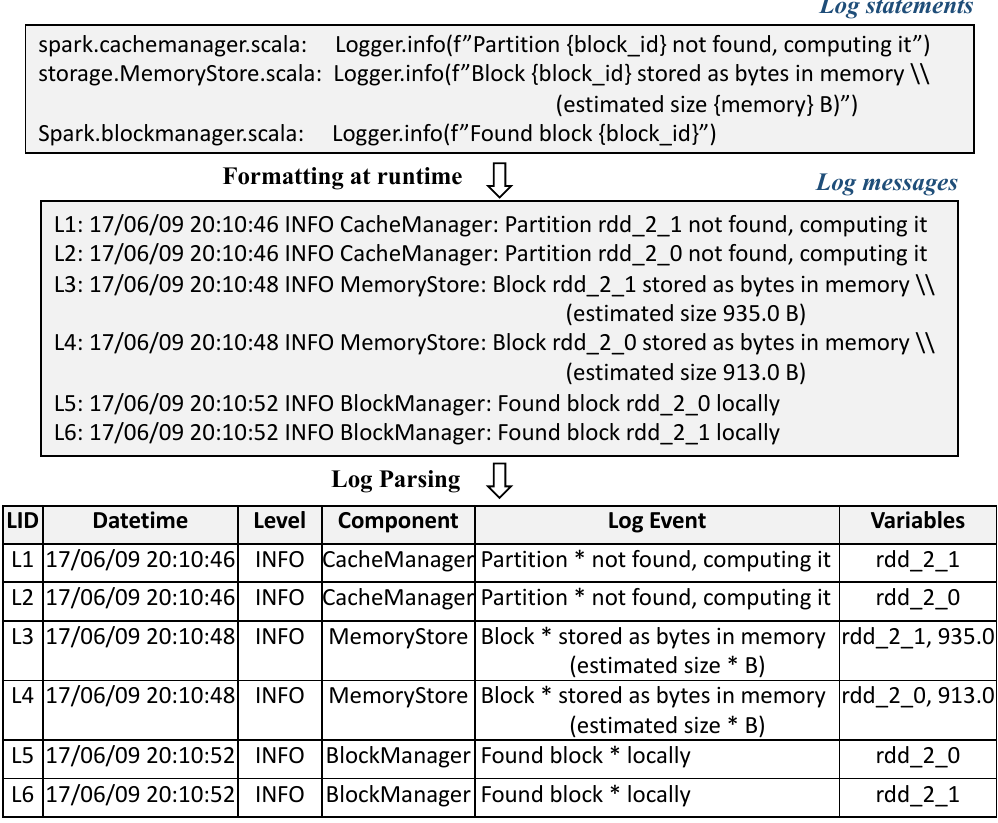}}
    \figin
    \caption{The generation and parsing of log data}
    \figin
    \label{fig:log_intro}
\end{figure}

Log data is essential for the maintenance and operation of software systems, which allows engineers to understand the system’s behaviors and diagnose problems~\cite{foalem2023studying,fu2014developers,yuan2012characterizing,he2018characterizing,li2020swisslog,li2022swisslog,li2022going,le2022log}.
Figure~\ref{fig:log_intro} illustrates the process of generating and parsing log messages. During the development phase, developers instrument log statements like \logtext{logger.info(``Partition \{block\_id\} not found, computing it'')} in source code~\cite{liu2019variables,yuan2012improving}. When the log statements are executed at runtime, a logger is specified to format the log statements to log content like \logtext{Partition rdd\_2\_1 not found, computing it}. Then the logger pads the log content to the log messages (e.g., \logtext{17/06/09 20:10:46 INFO CacheManager: Partition rdd\_2\_1 not found, computing it}) based on a pre-defined pattern including datetime, log level, and component information~\cite{yang2018nanolog, yao2018log4perf}. Log parsing~\cite{DBLP:conf/icse/LeZ23} is a fundamental step in log analysis that performs a reversible analysis to convert formatted log messages into structured ones. We introduce the structures of log parsing results as three components: \textit{headers} (e.g., datetime, log level, and component), \textit{log events} (e.g., \logtext{Partition * not found, computing it}) and \textit{variables} (e.g., \logtext{rdd\_2\_1}). A log sequence is a series of log messages, representing a system execution flow.

In practice, large-scale systems generate vast amounts of log data. These data must be stored for extended periods for a variety of reasons, including (i) Security analysis to identify potential long-term network attacks. For example, a recent attack remained undetected for eight months, and log access data is a critical data source to trace the attackers' behaviors~\cite{sanger2021scope, bing2020suspected}; (ii) Audit compliance. Cloud providers are required by local laws to store logs for a specified period; (iii) Long-term statistical analysis. Log data is analyzed for prolonged periods to gain insights into system performance and user behavior. However, archiving such a massive amount of log data introduces expensive overhead on log storage costs. 

Current logging frameworks such as ELK (ElasticSearch-LogStash-Kibana) stack~\cite{elk-stack} utilize general-purpose compressors (e.g., lzma~\cite{7za}, gzip~\cite{gzip}, bzip2~\cite{bzip2}) to compress a large volume of log data. These compressors analyze log files character-by-character, identify repeated characters in log data, and substitute them with a shorter representation. For instance, LZ77~\cite{ziv1977universal} algorithm used in gzip replaces the datetime \logtext{17/06/09 20:10:46} in L2 with a shorter representation of (\texttt{76}, \texttt{16}, \texttt{` '}), which comprises of an offset length, matching length, and the next character. 
However, general-purpose compressors do not consider the unique characteristics of log data. For example, the datetime here 
keeps increasing. 
Advanced log-specific compression methods leverage the latent structure of log data and apply various techniques to enhance the compression ratio of log data. For example, LogZip~\cite{liu2019logzip} extracts log events iteratively and replaces the log events with a shorter log event ID. LogReducer~\cite{wei2021feasibility} also parses log events and analyzes relations only in numerical parameter values. LogBlock~\cite{yao2021improving} targets applying some heuristics-based preprocessing steps on small log block compression. 
Although they achieve promising results on the log compression task, they either require some domain knowledge, such as manual datetime identification or apply many heuristics rules in preprocessing. Therefore, they still have limitations in practice.



\subsection{Related Work}


\textbf{General-purpose data compression methods.} Compression algorithms 
exploit statistical redundancy to eliminate redundancy literally, which can be categorized into three types: entropy-based, dictionary-based, and prediction-based. Entropy-based methods (e.g., Huffman encoding~\cite{huffman1952method}, Arithmetic encoding~\cite{witten1987arithmetic}) build a probability model and find the optimal minimized coding for data. Dictionary-based methods (e.g., gzip~\cite{gzip}, lzma~\cite{7za}) search the repeated tokens and replace them with dictionary references when processing the input stream. Prediction-based methods (e.g., PPMd~\cite{cleary1984data}, DeepZip~\cite{goyal2018deepzip}) use a set of previous tokens to predict the next token, and encode the prediction results adaptively. Yao et al.~\cite{yao2020study} conducted an empirical study on the performance of general-purpose compressors on log data. From the results, we can indicate that all general-purpose data compression methods can only analyze log messages character-by-character or bit-by-bit without reorganizing log data based on the characteristics of log data.

\textbf{Log-specific data compression methods.} Considering log data as semi-structured data, it is highly redundant by nature. Log-specific compression methods can be categorized into two types: non-parser-based and parser-based. Non-parser-based log compression methods (e.g., LogArchive~\cite{christensen2013adaptive}, Cowic~\cite{lin2015cowic} and MLC~\cite{feng2016mlc}) process log data without extracting log event patterns.
Parser-based log compression methods~\cite{liu2019logzip,ding2021elise,yao2021improving,wei2021feasibility} use a log parser to extract log events and process headers, events, and variables separately. According to the data type, they apply various encoding methods like delta encoding, common sub-pattern extraction, and dictionary encoding to shrink log data heuristically.
In terms of data characteristics, most of them~\cite{liu2019logzip, wei2021feasibility, yao2021improving} deal with the high redundancy of log content. Others~\cite{wei2021feasibility,yao2021improving,ding2021elise} leverage the timestamp feature. 
CLP~\cite{rodrigues2021clp} and LogGrep~\cite{wei2023loggrep} deliver efficient query tools on compressed data. CLP parses log lines into schemas and stores the variables as dictionary and non-dictionary variables. LogGrep further extracts static patterns and runtime patterns of dictionary variables and packs them into a set of capsules in a fine-grained manner. However, they could not achieve a high compression ratio while preserving searchable features.


%% file: sections/empirical-study.tex
\section{An Empirical Study on Characteristics of Log Data}
\label{sec:empirical}


In order to leverage the latent characteristics of log data to improve log compression, we conduct an empirical study on three real-world log datasets from a widely used log repository~\cite{loghubdatasets}. The basic statistics (i.e., file size, the number of log event types, and the number of log header fields) of the datasets are presented in Table~\ref{tab:empirical_study_ds}. Through the empirical study, we obtain three observations:
\input{tables/empirical_study_ds}


\textbf{Observation 1: Commonality and variability exist among pairs of log components.}
Considering a log sequence as a formatted execution flow, commonality and variability can be manifested in the following pairs. 

\begin{itemize}
    \item Header-Header (H-H): Each header is rendered by a pre-defined format, including timestamp, level and component. The timestamp exhibits a strong variability due to its increasing nature. The other meta information, such as log level and log component, shows a strong commonality because they usually come from a limited number of possible values.
    \item Event-Header (E-H): Among the padded information in log headers, there is some static information bound with events such as log level, logger name, and file location. For example, in Figure~\ref{fig:log_intro}, the log component of the event \texttt{``Partition * not found, computing it''} is always \texttt{CacheManager}. The commonality can be identified as the binding among them.
    \item Event-Variable (E-V): The number and the types of variables are always the same among log messages with the same log event, which can be considered as commonality. For example, in Figure~\ref{fig:log_intro}, the log messages L3 and L4 both have two variables. Their first variables are of the type of block ID (i.e., \texttt{``rdd\_x\_x''}) and their second variables are of the type of floating point number. 
    \item Variable-Variable (V-V): Variables within the same log event share a similar pattern with a slight difference. For example, in Figure~\ref{fig:log_intro}, variables in L1 and L2 are \texttt{``rdd\_2\_1''} and \texttt{``rdd\_2\_0''}, which follow the pattern \texttt{``rdd\_x\_x''}. Between them, the variability can be observed as the change of \texttt{x}, namely the change from 1 to 0. 
    
\end{itemize}
\input{tables/relation_empirical_study}


Two out of four pairs (i.e., H-H, E-H) can be explicitly defined in the empirical study.
For H-H pairs, if the header only includes a finite set possible set or an increasing timestamp, then it can be considered as a satisfied pair.
We identify E-H pairs as satisfying the condition if the log headers with the same log event only have one value.
We first calculate the total number of H-H and E-H pairs in three datasets accordingly.
Then we carefully identify the fields that meet the requirements shown above in H-H and E-H pairs.
For example, we count the 304 E-H possible pairs in total in Hadoop and identify 146 satisfied E-H pairs. The proportion of E-H pairs is calculated as 48\% (146/304).
The statistics of the above pairs occurrence and corresponding proportion results among three datasets are shown in Table~\ref{tab:relation_stat}. We can see from the results that the satisfied H-H pairs occupy 100\% among all headers. 
Satisfied E-H pairs also widely exist in log messages, which take around 29\% - 48\% of all E-H pairs.
The identified commonality and variability in these pairs can be leveraged to condense log data through dictionary replacement or differential values.




\begin{figure}[htbp]
    \figin
    \centerline{\includegraphics[width=0.48\textwidth]{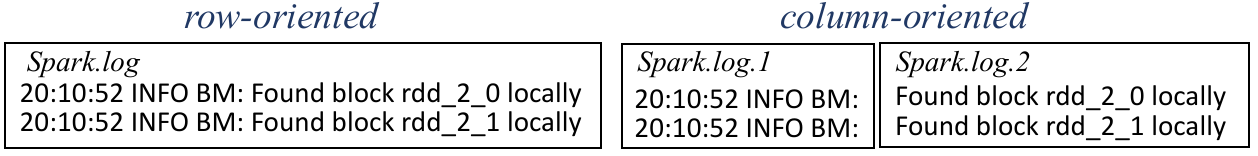}}
    \figin
    \caption{An example of different storage styles}
    \figin
    \label{fig:storage_style}
\end{figure}
\textbf{Observation 2: Storage style significantly affects the compression ratio.}
When reviewing the existing log-specific compression methods, we notice that they rarely consider the file storage style and explain the impact of file storage styles.
On the one hand, it is natural to store logs in a row-oriented storage style as log messages are printed line-by-line with logging frameworks~\cite{SCLTvaarandi2003data}. On the other hand, log files are well formatted as each log line is semi-structured text~\cite{ding2021elise}. Furthermore, there is evidence showing that general-purpose data compression methods perform better with structured or semi-structured texts~\cite{adiego2004lempel, nevill1996compressing} as much redundant information can appear in the structure.
Therefore, we argue that different storage styles can affect the performance of log compression.
We depict these two storage styles, i.e., column-oriented and row-oriented storage styles, in Figure~\ref{fig:storage_style}. The columns here are considered as fields of log messages, namely, log headers (e.g., \texttt{``20:10:52 INFO BM:''}) and log content (e.g., \texttt{``Found block rdd\_2\_0 locally''}) in Figure~\ref{fig:storage_style}.
Row-oriented storage style is to store columns of each log message in one row. Column-oriented storage style follows another way to store them column-by-column. For example, log headers are first stored in file 1 and then log content in file 2. 

We study the impact of different storage styles on log compression by analyzing three real-world log datasets. 
The original log files are stored in the row-oriented storage style.
Next, we transform logs to the column-oriented style by splitting each log line into its header and its content without further log parsing.
We then archive files in different storage styles using two widely-used general-purpose compression methods (i.e., lzma~\cite{7za} and gzip~\cite{gzip}).
We measure the size of the archived files and compute the improvement as $\Delta=\frac{row\ file\ size - column\ file\ size}{column\ file\ size}$. Table~\ref{tab:storagestyle} shows the results.

\input{tables/storage_style}

The results in Table~\ref{tab:storagestyle} demonstrate that utilizing a column-oriented storage style for log compression significantly reduces the compressed file sizes, compared to the row-oriented storage style.
On average, the column-oriented storage style achieves a storage reduction of 36\% to 103\% when using different general-purpose compression methods.
The main reason is that log data exhibits more common patterns 
with the column-oriented storage style than with the row-oriented style. 
This finding suggests that storing logs in a column-oriented manner could enable more effective compression.

\begin{figure}[bp]
\centering  
\subfigure[Hadoop]{
\label{fig:logpre}
\includegraphics[width=0.15\textwidth]{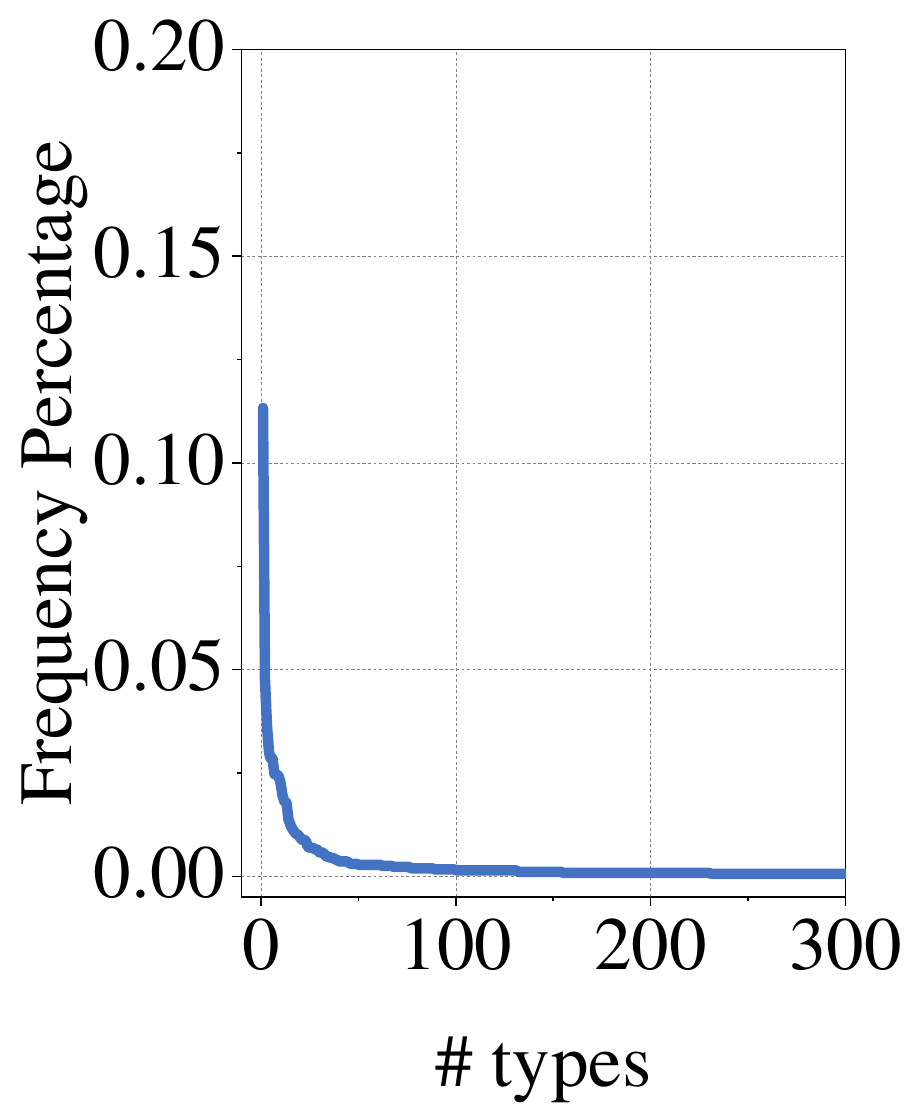}}
\subfigure[HPC]{
\label{fig:trace_compress}
\includegraphics[width=0.15\textwidth]{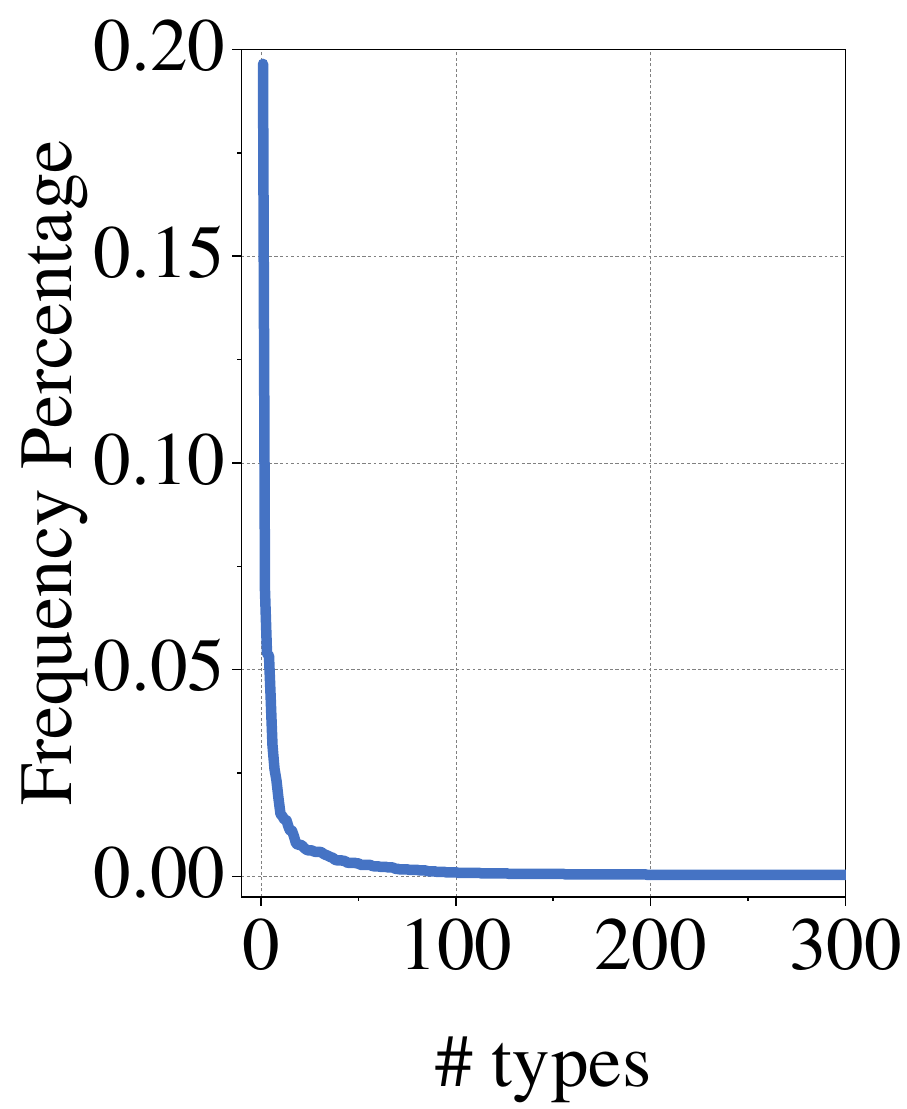}}
\subfigure[OpenSSH]{
\label{fig:trace_compress}
\includegraphics[width=0.15\textwidth]{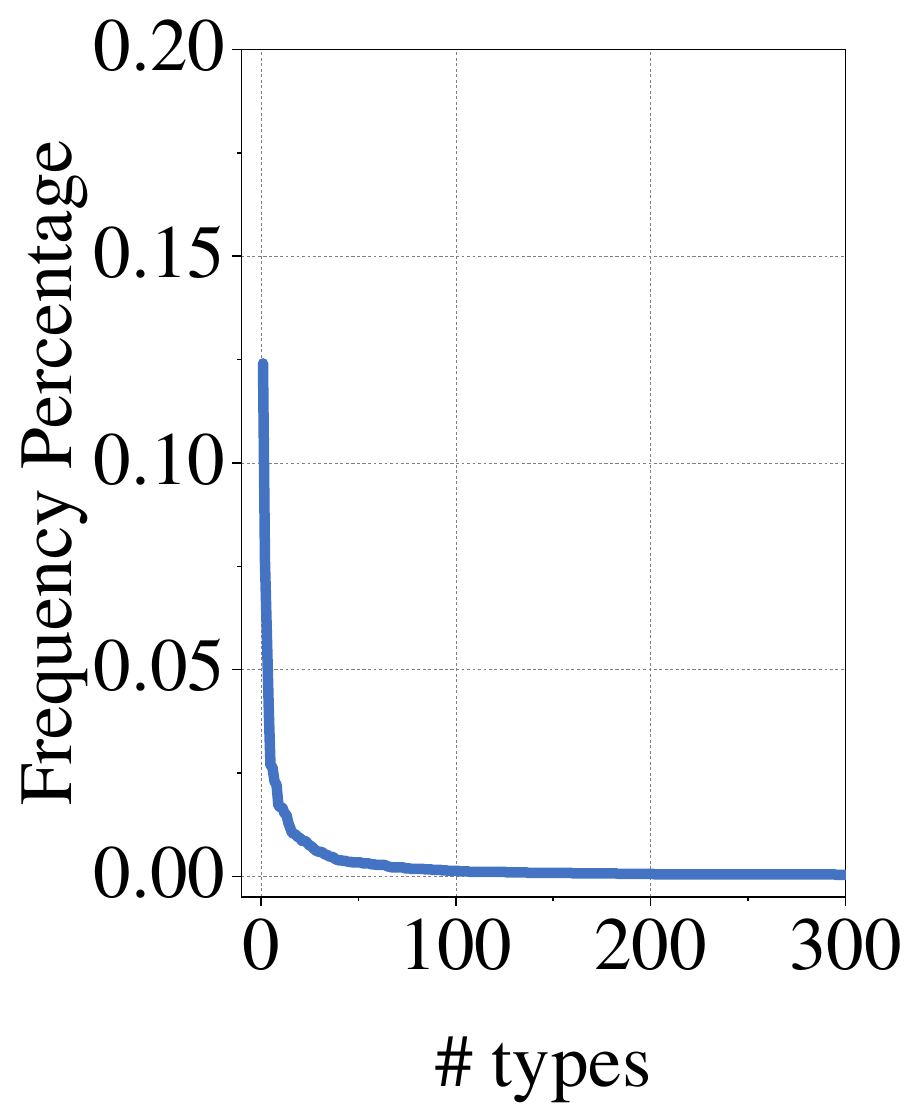}}
\figin
\caption{Imbalanced distribution of log sequence types}
\figin
\label{fig:log_seq_dist}
\end{figure}

\begin{figure*}[htbp]
    \centerline{\includegraphics[width=0.90\textwidth]{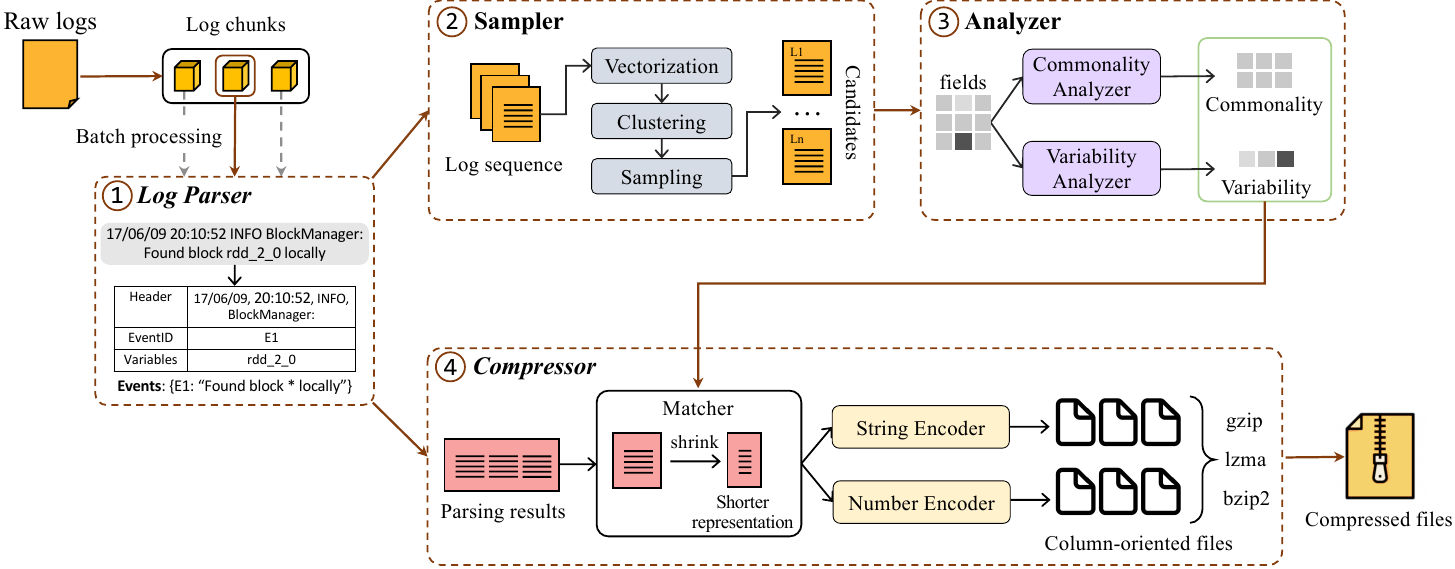}}
    \figin
    \caption{The overview of LogShrink framework}
    \figin
    \label{fig:framework}
\end{figure*}

\textbf{Observation 3: The distribution of log sequences is highly imbalanced.} 
A log sequence reflects an execution flow of a program~\cite{he2016experience}.
For example, in Figure~\ref{fig:log_intro}, three log events (i.e., \logtext{Partition * not found, computing it}, \logtext{Block * stored as bytes in memory (estimated size * B)}, and \logtext{Found block * locally}) form a type of log sequence. 
Analyzing the commonality and variability among all log sequences is time-consuming, thereby we study the distribution of log sequences to see if there are more efficient ways.
Prior work~\cite{he2018identifying} grouped log sequences by task ID and showed the long tail distribution of log sequence types. 
However, this method is not always feasible when 
log messages have no identifiers.
Instead of using identifiers, we follow recent studies~\cite{zhang2020anomaly, zhang2023semi} to group log sequences using a fixed-length window of size $h=50$, and consider two log sequences belonging to the same type if their similarity score is greater than a certain threshold (0.6 by default). The similarity score is calculated as the number of common tokens divided by the number of total tokens in two sequences. 
We then analyze the frequency percentage of each log sequence type in the three datasets and present the results in Figure~\ref{fig:log_seq_dist}, where the x-axis denotes the distinct log sequence types, and the y-axis represents the frequency proportion of each log sequence type in the dataset.
We can observe that the types of log sequences in the three datasets exhibit a highly imbalanced distribution, which is consistent across all three datasets. Out of hundreds of log sequence types, more than 50\% of the log sequences come from the first 3\%$\sim$5\% log sequence types.
The results suggest that we can identify commonality and variability more efficiently by analyzing a small sample of log sequences. 

%% file: tables/empirical_study_ds.tex
\begin{table}[htbp]
\small
\tabin
\caption{The statistics of datasets used in empirical study}
\tabin
\begin{center}
\begin{tabular}{c|rrr}
\toprule
\textbf{Dataset}   & \textbf{File size} & \textbf{\# Event} & \textbf{\# Header} \\ \midrule
Hadoop    & 48.61 MB  & 298      & 5         \\
HPC       & 32.00 MB  & 104 & 6         \\
OpenSSH   & 70.02 MB  & 62       & 5         \\
\bottomrule
\end{tabular}    
\end{center}
\tabin
\label{tab:empirical_study_ds}
\end{table}

%% file: tables/relation_empirical_study.tex
\begin{table}[htbp]
\small
\tabin
\caption{The statistics of occurrence and proportion of different pairs. \# denotes their occurrence and \% denotes their proportion.}
\tabin
\begin{center}
\begin{tabular}{c|cccccc}
\toprule
 \multirow{2}{*}{\textbf{Pairs}}  & \multicolumn{2}{l}{\textbf{Hadoop}} & \multicolumn{2}{l}{\textbf{HPC}} & \multicolumn{2}{l}{\textbf{OpenSSH}} \\
    & \#          & \%           & \#         & \%         & \#           & \%           \\ \midrule
H-H & 5           & 100\%            & 6          & 100\%          & 5            & 100\%            \\
E-H & 146         & 48\%         & 39         & 40\%       & 20           & 29\%        \\
\bottomrule
\end{tabular}    
\end{center}
\tabin
\label{tab:relation_stat}
\end{table}

%% file: tables/storage_style.tex
\begin{table}[htbp]
\small
\caption{Archived file size using different storage styles}
\tabin
\begin{center}
\resizebox{\linewidth}{!}{
\setlength{\tabcolsep}{4pt}
\renewcommand{\arraystretch}{1.1}
\begin{tabular}{c | c c c c c}
\toprule
\multirow{2}{*}{\textbf{Dataset}} & \multicolumn{2}{c}{\textbf{lzma}} & \multicolumn{2}{c}{\textbf{gzip}} & \multirow{2}{*}{Avg $\Delta$} \\ 
                         & Row      & Column ($\Delta$)    & Row     & Column ($\Delta$)     &                      \\ \midrule
Hadoop                   & 1.23 MB &	0.86 MB (0.43$\downarrow$) & 2.20 MB &	1.70 MB (0.29$\downarrow$) & 0.36$\downarrow$                 \\
HPC                      & 2.02 MB &	1.17 MB (0.72$\downarrow$) & 3.09 MB	&2.43 MB (0.27$\downarrow$) & 0.50$\downarrow$                 \\
OpenSSH                  & 3.92 MB &	1.73 MB (1.26$\downarrow$) & 4.19 MB &	2.34 MB (0.79$\downarrow$) & 1.03$\downarrow$         \\
\bottomrule
\end{tabular}   
}
\end{center}
\label{tab:storagestyle}
\end{table}

%% file: sections/methodology.tex
\section{LogShrink: The proposed Approach}
\label{sec:methodology}

Drawing upon the above observations, we propose LogShrink, a novel log compression method that can exploit the latent characteristics of log data to enable effective compression.
The overview of LogShrink framework is illustrated in Figure~\ref{fig:framework}. 
Since the raw log files that require compression are usually too large for processing, the raw log files are initially segmented into multiple log chunks with equal size. Log chunks are processed in batches, and each log chunk goes through four main components: \ding{172} \textit{Log Parser}, \ding{173} \textit{Sampler}, \ding{174} \textit{Analyzer}, and \ding{175} \textit{Compressor}.

The main essence behind \tool is that we try to represent log sequences in shorter forms based on their commonality and variability.
Therefore, in LogShrink, firstly, \textit{Log Parser} partitions log messages in each log chunk into three log components, namely log headers, log events, and log variables.
To exploit the commonality and variability observed in Observation 1, we propose a novel and effective 
\textit{analyzer} to identify the commonality and variability among log components. 
As it usually requires much time to analyze all log messages, a clustering-based sequence $sampler$ is introduced to accelerate the \textit{analyzer}. Based on Observation 3, the $sampler$ is designed to sample a small yet representative set of log sequences and then the $analyzer$ is applied to identify commonality and variability among sampled log sequences. The \textit{compressor} takes all the parsed results from \textit{log parser} and mined relations from \textit{analyzer} as input. It shrinks log data by replacing the log data with a shorter representation based on the analyzed commonality and variability. Then $encoders$ are applied according to value types (e.g., string values and numerical values).
Inspired by Observation 2, all the encoded data are stored in a column-oriented storage format and eventually compressed to log files by a general-purpose compressor such as lzma~\cite{7za} and gzip~\cite{gzip}. 



\subsection{Log Parser}

Log parsing is an essential step to convert unstructured log messages into structured ones. The procedure entails separating the log header and log content, followed by the identification of common and variable parts in the log content as log events and log variables, respectively. In recent years, many log parsers such as Spell~\cite{du2016spell}, LogMine~\cite{hamooni2016logmine}, Drain~\cite{he2017drain,he2018directed}, and LogPPT~\cite{DBLP:conf/icse/LeZ23} are proposed to achieve satisfactory effectiveness. However, these log parsers have a time complexity of approximately $O(n)$, rendering them impractical for log compression tasks, particularly for large log files. In this paper, we adopt a sub-optimal yet practical log parser proposed by LogReducer~\cite{wei2021feasibility}. It contains two steps: training and matching. Following this method, in the training step, the parser samples log segments and automatically generates header formats. Subsequently, it tokenizes the log segments and iteratively clusters them based on log level, component name, and frequently occurring words. After that, it builds a prefix tree to facilitate event matching where the first layer is the length of events, and the rest layer is tokens of events. In the matching step, the unsampled log messages utilize the built parser tree to search for the most similar event. If unmatched, it collects the raw unmatched log messages in an individual file. The log parser yields three log components, specifically header, events, and variables. Following Observation 2 in Section~\ref{sec:empirical}, we process all log components into column-oriented fields. Log headers are separated into multiple fields using space delimiters. We group log variables by the log event IDs, thereby the variable list with the same log event constructs a matrix. Then we could process and store all the log components in a column-oriented way.


\vspace{-6pt}
\subsection{Clustering-based Log Sequence Sampler}
\label{subsec:sampler}
\begin{figure}[htbp]
    \centerline{\includegraphics[width=0.47\textwidth]{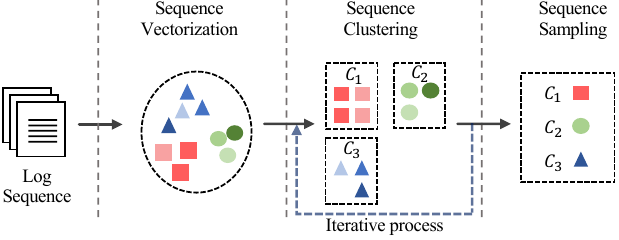}}
    \figin
    \caption{Clustering-based log sequence sampling}
    \figin
    \label{fig:seq_sampling}
\end{figure}

Based on Observation 3, the distribution of log sequence types is highly imbalanced, which brings a big challenge to sampling. To overcome this issue, we devise a clustering-based log sequence sampling to extract the representative log sequences. In this method, log messages are parsed into windows with a fixed length, denoted as $h$, to form a log sequence. The process of clustering-based log sequence sampling is depicted in Figure~\ref{fig:seq_sampling}.

\subsubsection{Sequence Vectorization}
We calculate vector representations for each log sequence. To determine the importance of each log event in log sequences, we adopt a widely used technique Inverse Document Frequency (IDF) in text mining. In this technique, log events that occur frequently across log sequences are assigned lower weights, while those that occur less frequently are assigned higher weights. Specifically, the IDF weight is defined as $w_{i d f}(e)=\log \left(\frac{N}{n_e}\right)$, where $N$ is the total number of log sequences and $n_e$ is the number of log sequences that contain log event $e$. We then construct sequence vectors using both log event frequency vectors $V_{fre}$ and log event weight vectors as Equation~\ref{eq:seq_vec}.

\begin{equation}
\label{eq:seq_vec}
 V_i = V_{fre} * [w_{idf}(e_1), ..., w_{idf}(e_n)]
\end{equation}

\subsubsection{Sequence Clustering}
We use an iterative clustering method to efficiently cluster log sequences. The process involves three steps: sampling, clustering, and matching. The input is a set of log sequences' vectors, and the output is their corresponding log sequence types. Specifically, given a set of $N$ log sequence vectors and a sample rate $\xi$ (default as $0.01$), we randomly select a subset of $M=min(\xi * N, k_{min})$ sequence vectors as the input for the $i$-th iteration. Here, $k_{min}$ is set to the minimum size that ensures the sampled data contains at least two samples of a single log sequence type. We calculate the value of $k_{min}$ as $k_{min}=\underset{k}{\mathrm{argmin}}\,(C^2_{k}p^{(k-2)}(1-p)^2 \ge 1-e^{-6})$. We then calculate the distance between each pair of sampled sequence vectors using Euclidean metrics (defined as $d(u,v)=\sqrt{\left\lVert u-v \right\rVert}$) and use a \textit{Hierarchical Agglomerative Clustering (HAC)} to cluster the sequences.
\textit{HAC} seeks to build a hierarchy of clusters in a bottom-up way. It performs a linkage between two clusters if their distance is smaller than a threshold $\theta$.
The resulting clusters yield $k$ sequence centers and the cluster IDs of all log sequences.

\subsubsection{Sequence Sampling}
Sequence sampling takes the candidate number $M$ as input and output $max(k, M)$ sampled log sequences. Given sampled data size $M$, we first calculate the sampled data size $m_{i} = \lceil M / k\rceil$ for each log sequence cluster. Then we randomly sample a data size of $m_{i}$ from each log sequence cluster and concatenate them. Finally, we obtain $max(k, M)$ log sequences. The impact under different window lengths, candidates $M$ and threshold $\theta$ are evaluated in Section~\ref{sec:evaluation}.

\vspace{-6pt}
\subsection{Commonality and Variability Analyzer}
\label{subsec:analyzer}
Observation 1 in Section~\ref{sec:empirical} indicates that \textit{commonality} and \textit{variability} widely exist in pairs of log components. 
They have the potential to generate a more concise representation of log data. Specifically, we define commonality as the common part among the log components and variability as the relatively steady change among the log components. 
For example, for a sequence $A1=\{$\texttt{task\#0-1}, \texttt{task\#1-2}, \texttt{task\#2-3}$\}$, the commonality is manifested as a common string pattern ``\texttt{task\#x-x}'', where $x$ denotes the variable parts among the sequence.
The variability in $A1$ is also obvious to observe. The task id is increasing by a step of 1 (i.e., from 0 to 1, and to 2). 
By exploiting such nature, we can present the corresponding log components by replacing the common part with a shorter index (e.g., \{\texttt{`task\#x-x'} $\rightarrow$ \texttt{0}\}) and the variable part with a shorter difference (e.g., 1), respectively.


\begin{figure}[htbp]
    \centerline{\includegraphics[width=0.47\textwidth]{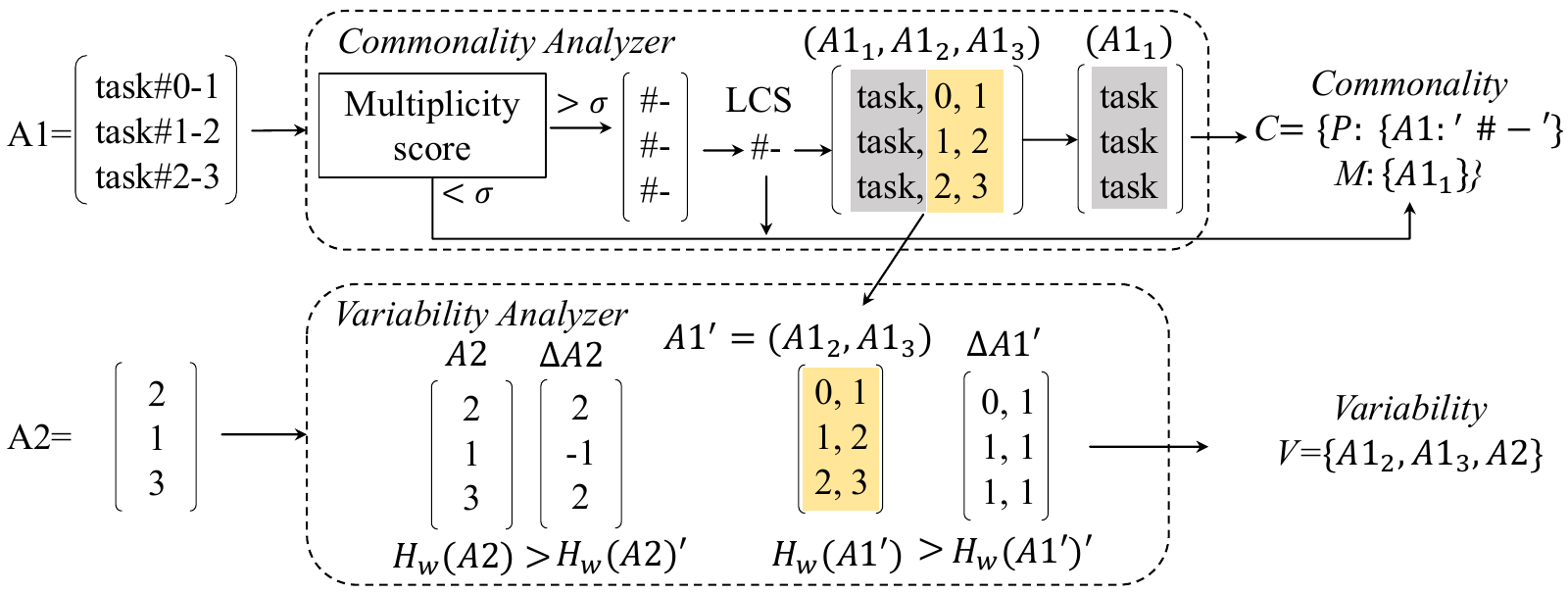}}
    \figin
    \caption{The process of analyzer}
    \figin
    \label{fig:relation_analyzer}
\end{figure}



Given a series of representative log sequences obtained from sequence sampler in Section~\ref{subsec:sampler}, Analyzer aims to analyze the latent commonality $C$ and variability $V$ in them. Figure~\ref{fig:relation_analyzer} demonstrates the process of Analyzer. In order to identify commonality in sequence, a multiplicity score is employed to measure whether a field contains multiple identical values. However, in cases such as $A1$, three values are distinct but share some common values which are separated by delimiters. To address this, we introduce a delimiter-level Longest Common Subsequence (LCS) to extract the common delimiters in the sequence (e.g.,  `\#-' in $A1$). This cannot be done by log parsers because they treat the entity with many delimiters inside (e.g., \texttt{rdd\_0\_1}, \texttt{/10.10.8.8}) as a single field. Consequently, $A1$ is divided into three sub-fields $A1_1, A1_2, A1_3$ using the mined delimiters. A fine-grained analysis is performed on these sub-fields. In the meantime, an entropy-based variability analyzer is introduced to see if the steps in the sequence follow a steady way. Given two input cases $A1$ and $A2$, the commonality $C$ of $A1$ is eventually concluded as the combination of the common delimiters and the field index as shown in Figure~\ref{fig:relation_analyzer}. The variability $V$ involves the indices of those satisfied fields.
Since the commonality we observed in log data mostly exists in string values in Section~\ref{sec:empirical}, we analyze the commonality in fields containing string values.
Similarly, we analyze the variability in fields containing numerical values.

\subsubsection{Commonality Analyzer}
Each field in log components usually shares the commonality in string pattern. We first use a multiplicity function to determine whether the field satisfies the multiplicity constraint. The multiplicity function is shown in Eq.~\ref{eq:sim_func}, where the multiset $X$ refers to the possible values of one field. We calculate the multiplicity by dividing the count of the unique set in $X$ by the count of the multiset $X$. For example, in Figure~\ref{fig:relation_analyzer}, given a multiset $A1$, the multiplicity score $M(A1) = \frac{3}{3}=1$. 
If the multiplicity score of one field is smaller than a preset threshold $\sigma$, we consider that the field is highly redundant so we can replace it with a shorter representation. Otherwise, LogShrink activates the LCS-based pattern miner to further analyze whether there is a fine-grained commonality in this field. 

\begin{equation}
\label{eq:sim_func}
M(X) = \frac{|\{x| x \in X\}|}{|X|}
\end{equation}

LCS is to find the longest common subsequence among a sequence set. Compared with token-level LCS used in other log parsers, we apply a delimiter-level LCS to mine the shared delimiters among the field.
Suppose $\Sigma$ is a universe of delimiters. Given any sequence $\alpha = \{a_1, a_2, ..., a_m\}$, we extract the delimiter sequence $d = \{d_1, d_2, ..., d_m\}$, such that $d_i \in \Sigma$. Then a delimiter subsequence of $d$ is defined as $\{d_i, d_{i+1}, ..., d_j\}$, where $i \in \mathbf{Z}^+$ and $1 \leq i \leq j \leq m$. A common subsequence is a subsequence of both sequences $d_i$ and $d_j$. In the case of log field $A1$, the delimiter sequences are extracted as $d=\{'\#-', '\#-', '\#-'\}$. The delimiter-level LCS of $d$ is `\#-'. 

Then, we separate each field by the mined common delimiters. The input example $A1$ is divided into three fields. We then calculate the multiplicity score of the first field and analyze the variability of the other two fields (i.e., numerical fields). Finally, the commonality of $A1$ is $C=\{P:\{A1:'\#-'\}, M:\{A1_1\}\}$, including the delimiter-level LCS and the field indices that satisfy the multiplicity constraint.

\subsubsection{Variability Analyzer}

The main characteristic of variability is that the data reveals a relatively stable differential. Given a sequence $\alpha = \{a_1, a_2, ..., a_m\}$, the difference of $\alpha$ is defined as $\Delta \alpha = \{a_1, a_2 - a_1, ..., a_m - a_{m-1}\}$. Entropy is a widely used measurement in information theory to measure disorder, randomness, and uncertainty. In this paper, we use entropy to measure the degree of variability. A higher entropy represents that the sequence contains more variability to store. The definition of entropy is given as follows:

\begin{equation}
    H(X)=\sum_{x \in X} P_X(x) \log_2 \frac{1}{P_X(x)}
\end{equation}
where, in the context of a field, we use $P(x)$ to represent the probability of observing the value $x$ and use $X$ to represent all of the possible values in this field. However, we also need to consider the output bits of these fields. Intuitively, small values take fewer bits of storage~\cite{wei2021feasibility,yang2018nanolog}. For example, a value of 3 only needs to take 2 bits to store. We use a metric $w(x)=\log_2(x+1)$ to measure the minimum required bits of storage. Then we incorporate the minimum required bits of storage with the entropy as a weighted entropy metric:

\begin{equation}
    \begin{aligned}
    H_w(X) &=\sum_{x \in X} w(x) P_X(x) \log_2 \frac{1}{P_X(x)} \\
        & = -\sum_{x \in X} \log_2(x+1) P_X(x) \log_2 {P_X(x)}
    \end{aligned}
\end{equation}

We calculate $H_w(X)$ of the original sequence $\alpha$ and the difference of sequence $\Delta \alpha$ as $H_w(\alpha)$ and $H_w(\Delta \alpha)$. If $H_w(\Delta \alpha)$ is smaller than $H_w(\alpha)$, the sequence $\alpha$ is considered to satisfy the variability, and vice versa. For example, in Figure~\ref{fig:relation_analyzer}, considering a sequence $A2=\{2,1,3\}$ and its difference $\Delta A2=\{2,-1,2\}$, the weighted entropy of them are 1.68 and 0.79, respectively. $A2$ is thus considered to satisfy the variability requirement. In summary, after analyzing the variability of given $A2$ and the two sub-fields $A1_2$, $A1_3$ in $A1$, the results of $V$ are that $V=\{A1_2, A1_3, A2\}$.

\subsection{Compressor}
\label{subsec:compressor}
The compressor takes all the parsing results from the log parser and identifies commonality and variability from the analyzer as input. It matches them to all log data and shrinks with a shorter representation. Then it encodes all values using an encoder according to its value types. All the encoded data are stored in multiple files in a column-oriented storage style. Finally, we use a general-purpose compressor as the zip tool to further improve the compression ratio. 

\subsubsection{Matcher}
After obtaining the identified commonality and variability from representative log sequences, we need to apply them to all log data. LogShrink shrinks log data by replacing the field's values with shorter representations according to their characteristic types.
For those fields whose commonality includes delimiter-level common patterns, we first separate them using the corresponding $P$. Then, for those fields in the commonality set satisfying the multiplicity constraint $M$,  we build a dictionary for repetitive tokens and replace these tokens with a dictionary index. The built dictionary and replaced data are stored as two files. As for variability, we perform the differencing operation between consecutive values for those fields in the variability set. For all the analyzed characteristics, if one of them cannot be satisfied in all log data, we will drop them and store the rest.


\subsubsection{Encoder}
We categorize objects in log data as two data types: string values and numerical values. For string values, LogShrink outputs their raw values. As the numerical values are fairly small, they only use a few bits specified by their types. For example, a 4-byte integer value of 13 only requires 4 bits to store losslessly. Following the method used in LogReducer~\cite{wei2021feasibility}, we adopt an elastic encoder to encode numerical values.

\subsubsection{Zip tool}

Based on Observation 2 in Section~\ref{sec:empirical}, the processed data are stored in a column manner. For example, each header (e.g., timestamp, log component) and each variable are stored in separate files to improve the compression ratio.
The final compression is done by a general-purpose compressor such as lzma~\cite{7za}, gzip~\cite{gzip}, and bzip2~\cite{bzip2}. 


\subsection{Decompressor}
\label{subsec:decompressor}
The decompression process is a reverse process of compression. At first, the decompressor applies the general-purpose decompressor (e.g., lzma~\cite{7za}, gzip~\cite{gzip}, bzip2~\cite{bzip2}) to decompress the whole compressed file into many uncompressed files. Next, LogShrink loads the files that record the identified commonality and variability sets. Then, LogShrink performs the recovery process of differencing operations on fields and the dictionary mapping operation accordingly. After that, we obtain three log components which are generated by the log parser. To recover the message content, we parameterize the asterisk in log events using variables in order. Finally, the header and message content can be joined together with the extracted delimiters to obtain the original log messages.

%% file: sections/evaluation.tex
\section{Evaluation}
\label{sec:evaluation}
We conduct extensive experiments on a variety of log datasets. Our evaluation aims to answer three questions: 
\begin{itemize}
    \item RQ1: What is the overall performance of LogShrink?
    \item RQ2: What is the effect of each individual component in LogShrink?
    \item RQ3: What is the impact of different settings?
\end{itemize}

\subsection{Experimental Design}

\subsubsection{Datasets}
We use 16 representative log datasets~\cite{loghubdatasets} from a wide range of systems to evaluate LogShrink, including distributed systems (e.g., HDFS, Hadoop, Spark, Zookeeper, OpenStack),  supercomputers (e.g., BGL, HPC, Thunderbird), operating systems (e.g., Windows, Linux, Mac), mobile systems (e.g., Android, HealthApp), server applications (e.g., Apache, OpenSSH), and standalone softwares (e.g., Proxifier).
All these logs amount to over 77 GB in total. The details of experiment datasets are presented in Table~\ref{tab:dataset_stat}.

\input{tables/dataset_statistics}

\subsubsection{Evaluation Metrics}
To measure the performance of LogShrink in log compression, we use the compression ratio and compression speed, which are widely used in the evaluation of compression methods~\cite{liu2019logzip,wei2021feasibility,yao2021improving}. The definitions are given as follows:
\begin{equation}
\label{eq:cr}
    \text{Compression Ratio } (CR)=\frac{Original\ File\ Size}{Compressed\ File\ Size}
\end{equation}

\begin{equation}
\label{eq:cs}
    \text{Compression Speed } (CS)=\frac{Original\ File\ Size}{Cost\ Time}
\end{equation}

\subsubsection{Baselines}
We compare our proposed method with two state-of-the-art log compression methods (e.g., LogZip~\cite{liu2019logzip} and LogReducer~\cite{wei2021feasibility}) and three representative general-purpose compression methods (e.g., gzip~\cite{gzip, deutsch1996gzip}, lzma~\cite{7za}, and bzip2~\cite{bzip2}). gzip~\cite{gzip} is a traditional compression method based on DEFLATE algorithm, which achieves a good compression speed instead of a good compression ratio. Compared to gzip, bzip2~\cite{bzip2} is based on the Burrows-Wheeler transform algorithm. It has a better compression ratio yet worse compression speed. Lzma~\cite{7za}, which uses dictionary compression algorithms, usually gets a better compression ratio but has a relatively slow compression speed. In this study, we use a python package \textit{tarfile}~\cite{tarfile} to compress data with gzip and bzip2. The compression level opts to the highest 9 to achieve the best compression ratio. For lzma, we use the standalone \textit{7za} tool~\cite{7za} in the Linux system to compress data. In terms of state-of-the-art log-specific compression methods, LogZip extracts hidden structures for all log messages and performs different levels of compression.
We select the level 3 to achieve the highest compression ratio.
LogReducer is also a log parser-based compressor, which can further compress numerical values. 
We use their open-sourced code~\cite{lgozipcode, logreducercode} in our experiments. 

\subsubsection{Implementation and Environment}
We implement LogShrink in Python 3.8. 
The raw log files are segmented with an equal size of 100k lines. We adopt the log parser implemented by the LogReducer~\cite{wei2021feasibility} and modify it to adapt to our framework. As for the setting of parameters, we pre-define the universe of delimiters $\Sigma=\{-\#><\_:;,[]\backslash/.()\}$ used in the commonality analyzer. We set threshold $\theta=4$, fixed-window length $h=20$, and $M=16$ used in sequence sampler as defaults. The threshold $\sigma$ used in the commonality analyzer is set to $0.3$ by default. Since the sampling in LogReducer~\cite{wei2021feasibility} and LogShrink might yield random results during execution, we run them 10 times for all experiments and obtain the average results. We conduct our experiments on a Linux server equipped with 8$\times$ Intel Xeon 2.2GHz CPUs (with 32 cores in total) and 128GB RAM, and Red Hat 8.1 with Linux kernel 4.18.0.

\subsection{RQ1: The Overall Performance of LogShrink}

In this RQ, we compare \tool with state-of-the-art tools for log compression, including three general-purpose compressors (i.e., gzip~\cite{gzip}, lzma~\cite{7za}, and bzip2~\cite{bzip2}) and two log-specific compressors (i.e., LogZip~\cite{liu2019logzip} and LogReducer~\cite{wei2021feasibility}). 

\subsubsection{Effectiveness:}
Firstly, we compare the results of \tool with baselines in terms of Compression Ratio (CR). For a fair comparison, we use lzma from the 7z packet~\cite{7za} as the zip tool for both LogZip and LogReducer. Table~\ref{tab:r1_cr} shows the results.

\input{tables/rq1_cr}

From the results, we can see that \tool outperforms existing methods or achieves comparable results on almost all datasets (14 out of 16 datasets) in terms of CR. Specifically, \tool exceeds all general-purpose compressors. It can achieve a CR of 4.57$\times$ on average and 25.64$\times$ at best over gzip, a traditional compression algorithm. Besides, \tool achieves 1.16$\times$ to 5.54$\times$ CR compared to lzma, and 1.14$\times$ to 6.76$\times$ compared to bzip2. In the comparison with two log-specific compression methods, \tool significantly outperforms LogZip by achieving better CRs on 15 out of 16 datasets. It achieves a CR of 1.66$\times$ on average and 2.87$\times$ at best (on Spark) over LogZip.
Moreover, \tool equips higher CR on 15 out of 16 datasets compared to the most powerful log compressor, LogReducer.
In particular, it exceeds LogReducer by 4.56\% (Spark) to 33.04\% (Windows). On the Thunderbird dataset, \tool also performs comparably by achieving a high CR of 98.47\% over LogReducer.
It is worth noting that on large-scale datasets (i.e., BGL, HDFS, Spark, and Windows) except Thunderbird, \tool performs the best compared to other log-specific compression methods. 
Its CR is 1.05$\times$ to 2.87$\times$ that of LogZip and LogReducer.
Note that, in our experiments, LogZip failed to parse and compress Thunderbird data within 1 week. 
The reason why LogShrink performs worse on Android dataset than LogZip is that we adopt a sub-optimal yet practical log parser in our work but Logzip adopts an optimal yet slower log parser. The number of templates in Android is up to 76,923, making it ineffective in extracting log events from a limited number of samples. 

\subsubsection{Efficiency:} Our \tool explicitly aims at compressing log files with high compression ratio in a reasonable running time. Therefore, we next analyze and compare \tool with log-specific compression methods in terms of Compression Speed.
Table~\ref{tab:rq1-cs} shows the results.

\input{tables/rq1_cs}

We can see that \tool can compress log files with reasonable efficiency. It can achieve a compression speed of 2.95 MB/s on average, ranging from 1.31 to 5.51 MB/s. Compared to LogZip, which is also a Python-based compressor, \tool outperforms it significantly in efficiency. Specifically, \tool is 1.83$\times$-273.79$\times$ (26.6$\times$ on average) as fast as LogZip. \tool is generally slower than LogReducer, as LogReducer is written in C++ and thus is more optimized in terms of execution speed than LogShrink (written in Python). 


\subsection{RQ2: Ablation Study}

\input{tables/rq2_ablation_cr}

To evaluate the effectiveness of individual components in LogShrink, we perform an ablation study among the full LogShrink, LogShrink without clustering-based sequence sampling (denoted as w/o Sampler), and LogShrink without commonality and variability analyzer (denoted as w/o Analyzer). The experiment results of both compression ratio and speed are presented in Table~\ref{tab:r2_cr}.


The contribution of commonality and variability analyzer can be observed in the comparison between LogShrink and LogShrink w/o Analyzer.
We can see from the results that LogShrink can achieve an improvement of 23.12\% in CR on average.
Specifically, 
LogShrink achieves an improvement ranging from 2.92\% (on Proxifier) to 64.78\% (on Zookeeper) in terms of CR compared to LogShrink without Analyzer. 
The improvement largely depends on the amount of commonality and variability in log data.
It is worth noting that LogShrink w/o Analyzer also stores the log parsing results using the column-oriented format. From the results, we can see that without Analyzer, LogShrink can achieve 1.1 $\times$ to 3.8 $\times$ CR compared to lzma, which is consistent with Observation 2 in Section~\ref{sec:empirical}.
In terms of CS, LogShrink performs 0.64 $\times$ slower than LogShrink w/o Analyzer. It is a trade-off to consider both CR and CS in practical usage. The CS with a higher CR is probably slower than the CS with a lower CR. 

The clustering-based sequence sampling contributes to the improvement of CS.  
In Table~\ref{tab:r2_cr}, we can observe that LogShrink is faster than the LogShrink w/o Sampler by 8.03\% to 42.52\%, which is consistent with the expectation. 
In terms of CR, LogShrink achieves comparable results on 13 out of 16 datasets although we perform the commonality and variability analysis on a small number of representative log sequences out of all log sequences. However, not all datasets can mine the representative log datasets effectively. LogShrink shows a slight decrease of 10.0\% to 14.7\% in dataset Windows and HDFS. This is because the number of log sequence types in Windows and HDFS are much higher than others, their CR will be significantly impacted by the parameters set in clustering-based sequence sampling.

\subsection{RQ3: The Impact of Different Settings}
In this RQ, we explore the impact of critical parameters (i.e., the window length $h$ to form a log sequence, the number of sampling samples $M$, the distance threshold $\theta$, and the zip tool) on LogShrink. Due to the space constraint, we show the evaluation results on 3 out of 16 datasets (i.e., BGL, HDFS, and Spark). 

\begin{figure*}[t]
\centering  
\subfigure[Impact of window length $h$]{
\label{fig:rq3-wh}
\includegraphics[width=0.3\textwidth]{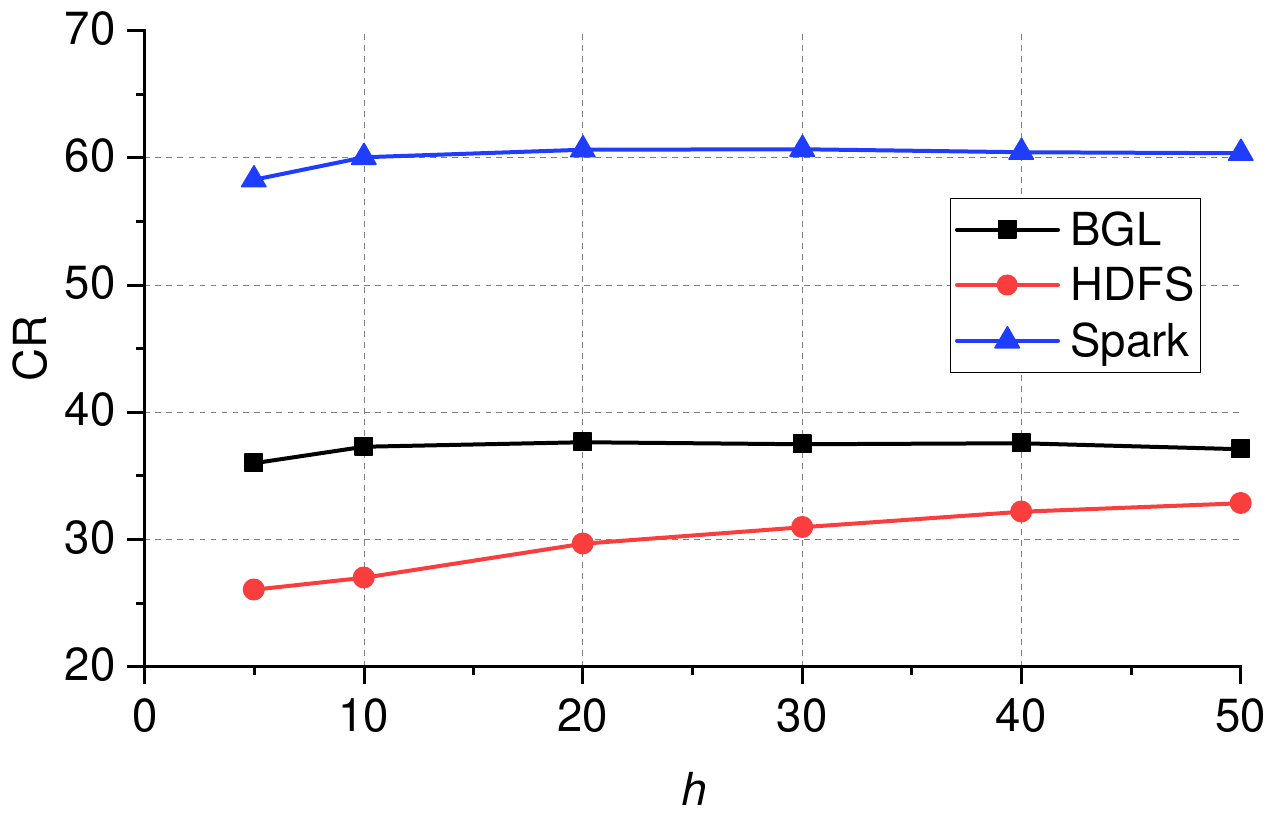}}
\subfigure[Impact of sampling samples $M$]{
\label{fig:rq3-nc}
\includegraphics[width=0.3\textwidth]{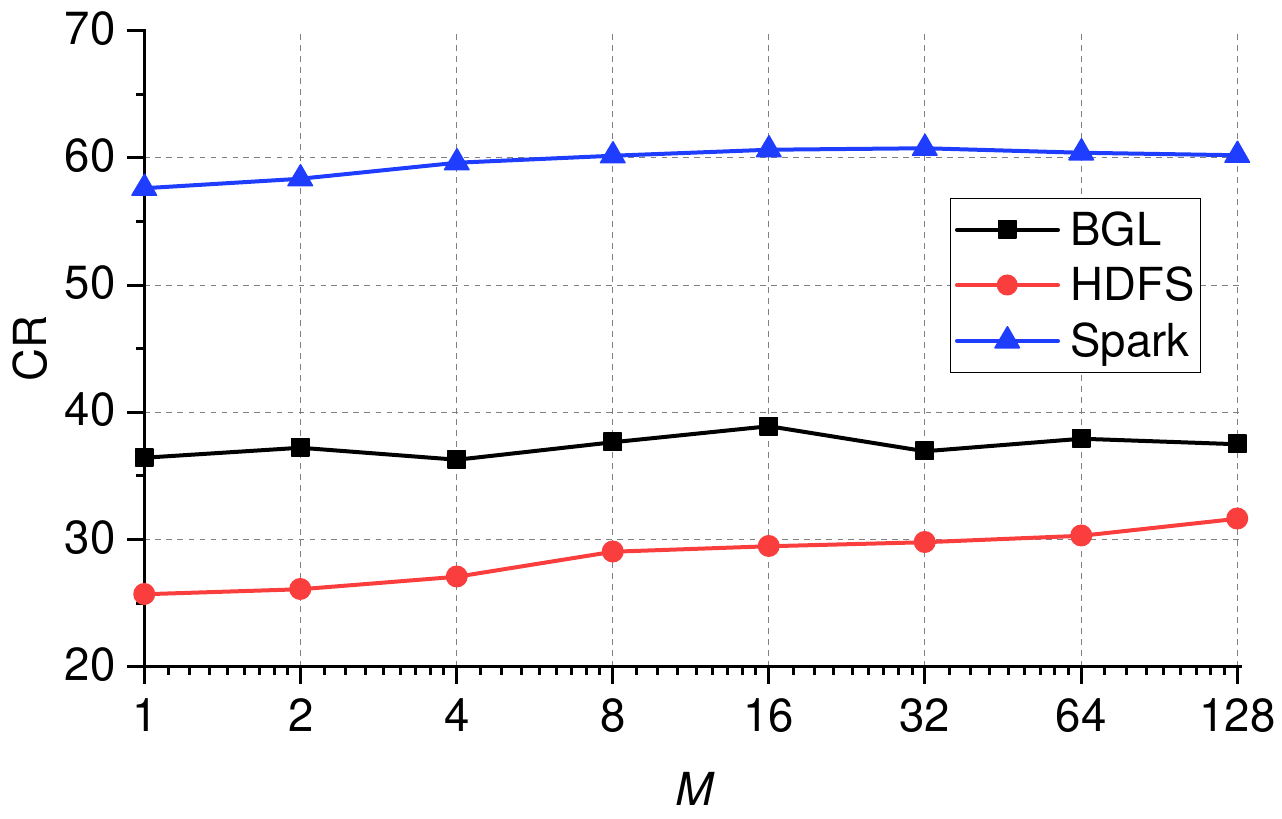}}
\subfigure[Impact of distance threshold $\theta$]{
\label{fig:rq3-th}
\includegraphics[width=0.3\textwidth]{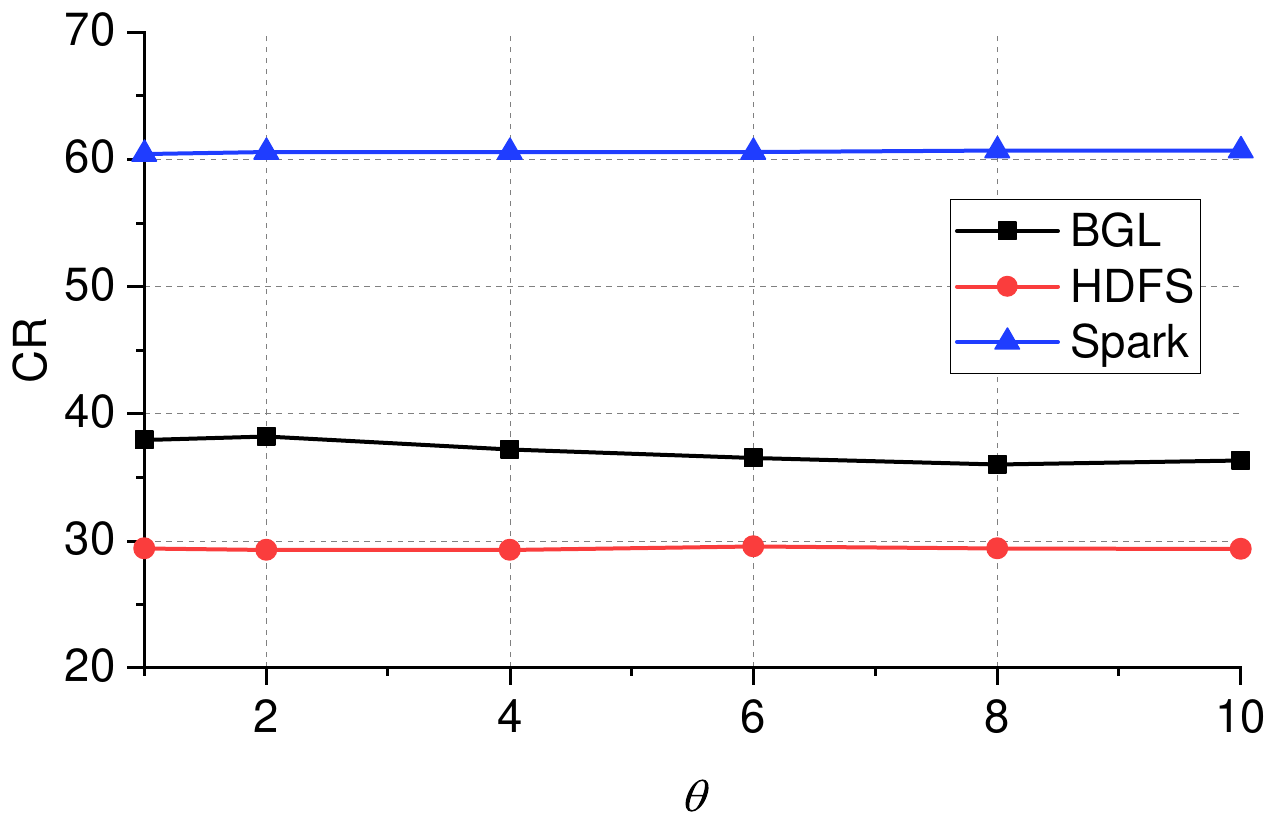}}
\figin
\caption{Impact of different settings in clustering-based sequence sampling}
\figin
\label{fig:rq3-sq}
\end{figure*}

\textbf{Impact of windows length $h$ in sequence sampling.} A larger window length $h$ can involve more log messages and generate more log sequence types. We run LogShrink with different window lengths $h$ in the range [5,50]. Figure~\ref{fig:rq3-wh} presents the experimental results. We can observe that the CR in all datasets has a sheer increase when $h$ changes from 5 to 10. But for other $h$ in [10,50], all datasets except HDFS show relatively stable results in CR. Hence, we select $h=20$ as the default, which shows a relatively high CR.

\begin{figure*}[t]
\centering  
\subfigure[gzip]{
\label{fig:rq3-k-gzip}
\includegraphics[width=0.3\textwidth]{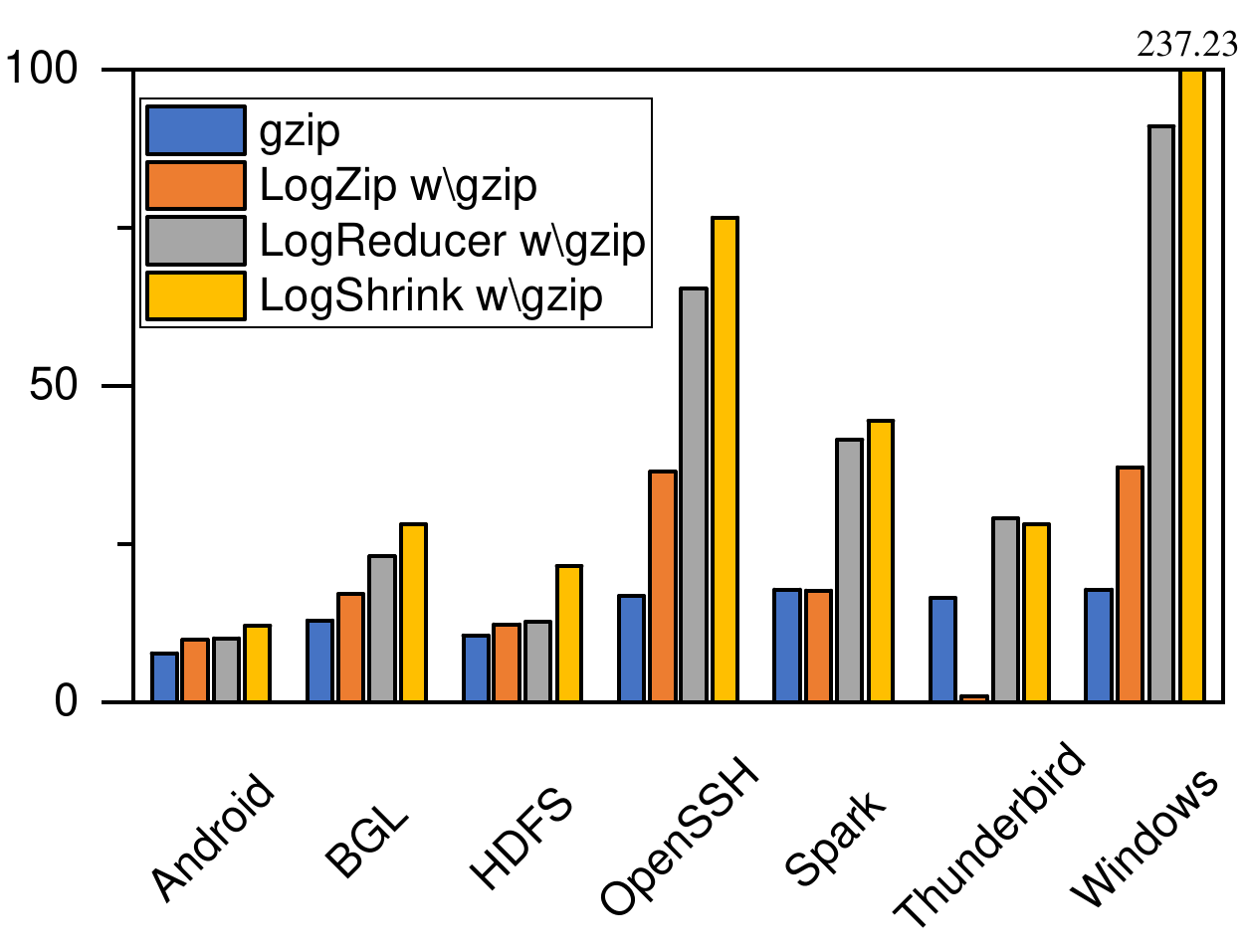}}
\subfigure[bzip2]{
\label{fig:rq3-k-bzip2}
\includegraphics[width=0.3\textwidth]{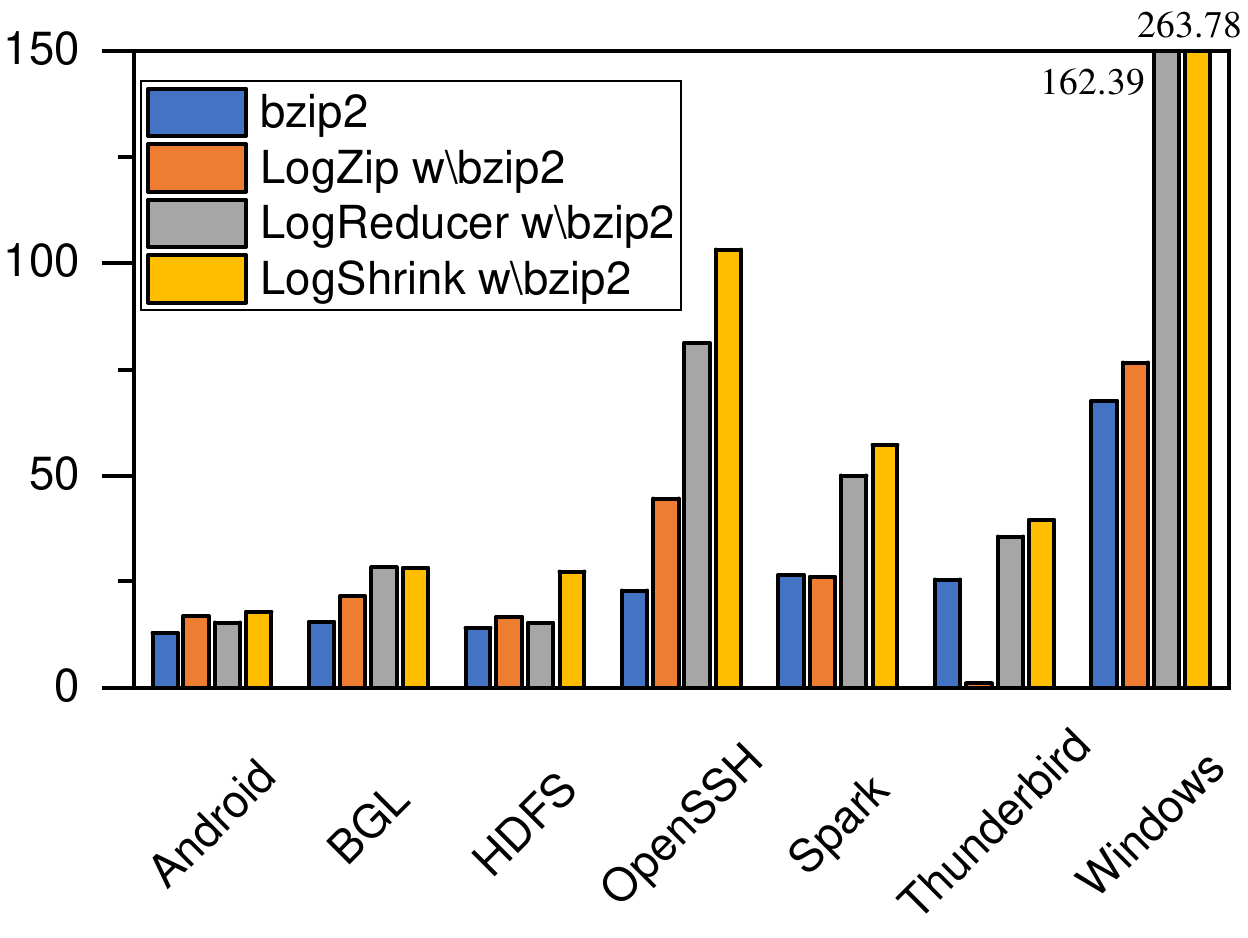}}
\subfigure[lzma]{
\label{fig:rq3-k-lzma}
\includegraphics[width=0.3\textwidth]{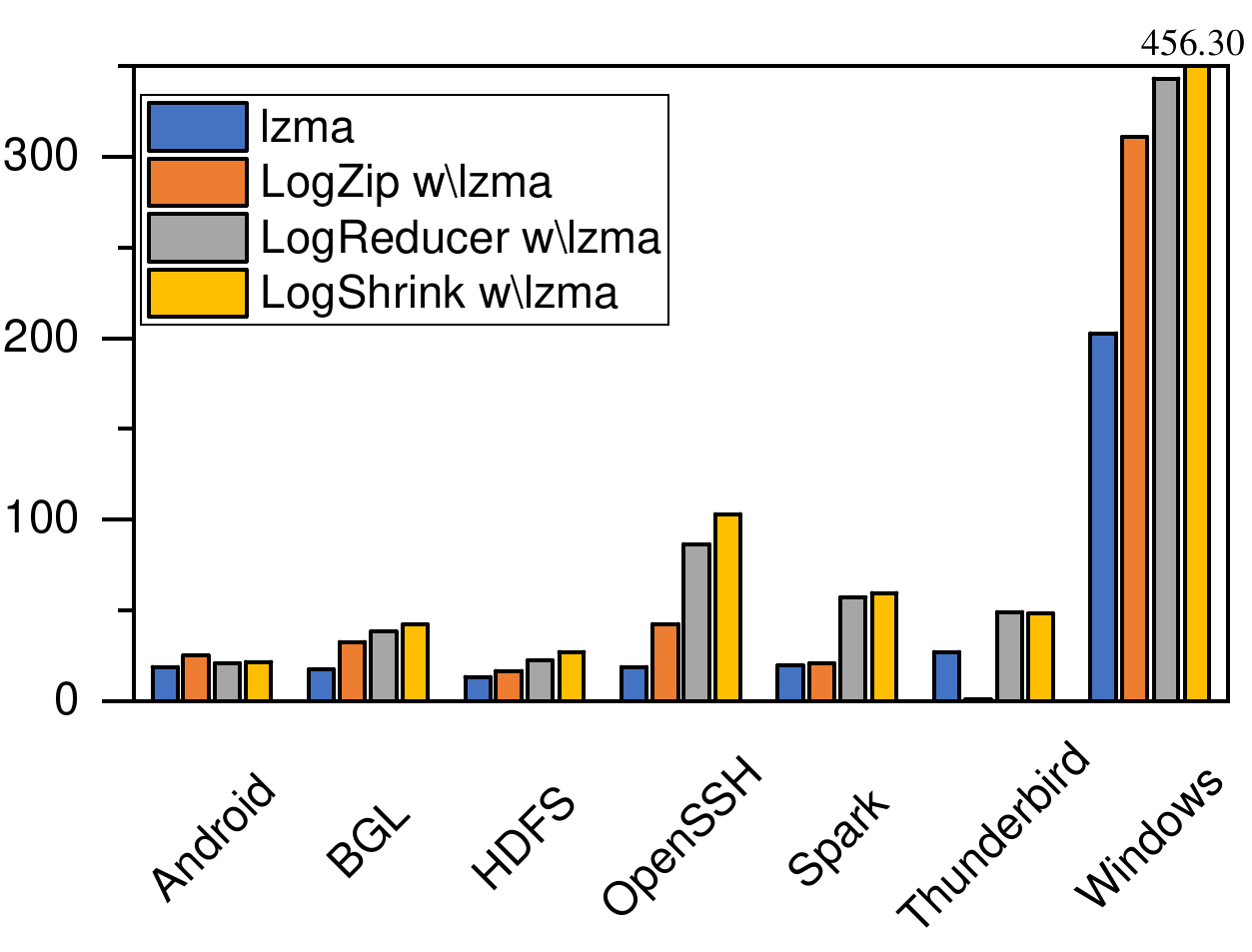}}
\figin
\caption{Impact of different zip tools used in compressor}
\figin
\label{fig:rq3-k}
\end{figure*}

\textbf{Impact of $M$ sampled candidates in sequence sampling.} 
A higher $M$ means that more log sequences can be fed into the analyzer.
We adjust $M$ from $2^0$ to $2^7$ and observe the changes in CR among the three datasets. The experiment results are illustrated in Figure~\ref{fig:rq3-nc}. From the figure, we can see that the CR increases with rising $M$. Especially, the CR of HDFS is significantly affected by the parameter $M$, showing a 23.1\% increase in $M=128$ compared to $M=1$. From the results, we select $M=16$ as the default in \tool, considering the trade-off between CR and CS.

\textbf{Impact of the distance threshold $\sigma$ in sequence sampling.} A larger distance threshold $\theta$ represents that the coverage of one cluster is wider and 
the number of clusters is less.
Figure~\ref{fig:rq3-th} shows the experimental results of the distance threshold $\theta$ in the range of [1,10]. The CR of BGL shows a steady decline with $\theta$ rising in the range [2,10]. However, the CR of the other two datasets remains stable as $\theta$ grows. From the results, we can see that the distance threshold $\theta$ insignificantly affects the CR, thereby it is set to 4 by default.

\textbf{Impact of different zip tools used in compressor.} LogShrink's CR is influenced by the zip tools employed in the compressor. To demonstrate this, we experimented with different zip tools (i.e., gzip, bzip2, lzma) on LogShrink and two other state-of-the-art methods (i.e., LogReducer, LogZip). Due to the space limit, we present the evaluation results for the top 7 largest datasets in Figure~\ref{fig:rq3-k}. The consistent superiority of LogShrink is observed across different zip tools. In comparison to the best CR performance among all baselines, LogShrink exhibits an improvement of 15.07\%, 30.67\%, and 15.30\% on average with gzip, bzip2, and lzma, respectively.



%% file: tables/dataset_statistics.tex
\begin{table}[htbp]
\caption{The statistics of experimental datasets}
\tabin
\begin{center}
\resizebox{.93\linewidth}{!}{
\setlength{\tabcolsep}{4pt}
\renewcommand{\arraystretch}{1.1}
\begin{tabular}{c|crr}
\toprule
\textbf{System Type }                        & \textbf{Dataset} & \textbf{File Size} & \textbf{\# Lines }    \\ \midrule
\multirow{5}{*}{Distributed systems} & HDFS            & 1.47  GB       & 11,175,629  \\
                                     & Hadoop          & 48.61  MB      & 394,308     \\
                                     & Spark           & 2.75 GB       & 33,236,604  \\
                                     & Zookeeper       & 9.95 MB       & 74,380      \\
                                     & OpenStack       & 58.61 MB      & 207,820     \\ \hline
\multirow{3}{*}{Supercomputers}      & BGL             & 708.76 MB     & 4,747,963   \\
                                     & HPC             & 32.00 MB      & 433,489     \\
                                     & Thunderbird     & 29.60 GB      & 211,212,192 \\ \hline
\multirow{3}{*}{Operating systems}   & Windows         & 26.09 GB      & 114,608,388 \\
                                     & Linux           & 2.25 MB       & 25,567      \\
                                     & Mac             & 16.09 MB      & 117,283     \\ \hline
\multirow{2}{*}{Mobile systems}      & Android         & 183.37 MB     & 1,555,005   \\
                                     & HealthApp       & 22.44 MB      & 253,395     \\ \hline
\multirow{2}{*}{Server applications} & Apache          & 4.90 MB       & 56,481      \\
                                     & OpenSSH         & 70.02 MB      & 655,146     \\ \hline
Standalone software                  & Proxifier       & 2.42 MB       & 21,329     \\
\bottomrule
\end{tabular}    
}
\end{center}
\tabin
\label{tab:dataset_stat}
\end{table}

%% file: tables/rq1_cr.tex
\begin{table}[h]
\caption{Comparison in terms of Compression Ratio}
\tabin
\label{tab:r1_cr}
\centering
\resizebox{\linewidth}{!}{
\setlength{\tabcolsep}{1.5pt}
\renewcommand{\arraystretch}{1.2}
\begin{tabular}{ccrcccc}
\hline
\textbf{Dataset}   & \textbf{gzip} & \textbf{lzma} & \textbf{bzip2} & \textbf{LogZip} & \textbf{LogReducer} & \textbf{LogShrink} \\ \hline
Android     & 7.742         & 18.857        & 12.787         & \textbf{25.165} & 20.776              & 21.857             \\
Apache      & 21.308        & 25.186        & 29.557         & 30.375          & 43.028              & \textbf{55.940}    \\
BGL         & 12.927        & 17.637        & 15.461         & 32.655          & 38.600              & \textbf{42.385}    \\
Hadoop      & 20.485        & 36.095        & 32.598         & 35.008          & 52.830              & \textbf{60.091}    \\
HDFS        & 10.636        & 13.559        & 14.059         & 16.666          & 22.634              & \textbf{27.319}    \\
HealthApp   & 10.957        & 13.431        & 13.843         & 22.632          & 31.694              & \textbf{39.072}    \\
HPC         & 11.263        & 15.076        & 12.756         & 27.208          & 32.070              & \textbf{35.878}    \\
Linux       & 11.232        & 16.677        & 14.695         & 23.368          & 25.213              & \textbf{29.252}    \\
Mac         & 11.733        & 22.159        & 18.074         & 26.306          & 35.251              & \textbf{39.860}    \\
OpenSSH     & 16.828        & 18.918        & 22.865         & 42.606          & 86.699              & \textbf{103.175}   \\
OpenStack   & 12.158        & 14.437        & 15.231         & 17.258          & 16.701              & \textbf{22.157}    \\
Proxifier   & 15.716        & 18.982        & 23.619         & 21.493          & 25.501              & \textbf{27.029}    \\
Spark       & 17.825        & 19.908        & 26.497         & 20.825          & 57.135              & \textbf{59.739}    \\
Thunderbird & 16.462        & 27.309        & 25.428         & —               & \textbf{49.185}     & 48.434             \\
Windows     & 17.798        & 202.568       & 67.533         & 310.596         & 342.975             & \textbf{456.301}   \\
Zookeeper   & 25.979        & 27.667        & 36.156         & 47.373          & 94.562              & \textbf{116.981}   \\ \hline
\multicolumn{7}{l}{%
  \begin{minipage}{\linewidth}%
    \vspace{3pt}
    Note: ``---" denotes timeout. LogZip cannot parse and compress Thunderbird log within 1 week.
  \end{minipage}%
}\\
\end{tabular}
}
\end{table}

%% file: tables/rq1_cs.tex
\begin{table}[htbp]
\centering
\caption{Comparison in terms of Compression Speed (MB/s)}
\vspace{-6pt}
\label{tab:rq1-cs}
\setlength{\tabcolsep}{3pt}
\renewcommand{\arraystretch}{1.1}
\resizebox{.7\linewidth}{!}{
\begin{tabular}{@{}cccc@{}}
\toprule
     \textbf{Dataset}       & \textbf{LogZip} & \textbf{LogReducer} & \textbf{LogShrink} \\
            \midrule
Android     & 0.068           & 8.918               & 5.347              \\
Apache      & 0.737           & 1.686               & 1.880              \\
BGL         & 0.874           & 18.189              & 2.519              \\
Hadoop      & 0.901           & 4.919               & 3.137              \\
HDFS        & 0.701           & 20.570              & 3.253              \\
HealthApp   & 0.736           & 4.108               & 2.064              \\
HPC         & 0.644           & 5.110               & 2.485              \\
Linux       & 0.687           & 0.526               & 1.307              \\
Mac         & 0.009           & 2.887               & 2.572              \\
OpenSSH     & 0.715           & 13.268              & 3.409              \\
OpenStack   & 0.537           & 6.389               & 2.945              \\
Proxifier   & 0.716           & 0.929               & 1.315              \\
Spark       & 0.550           & 18.871              & 2.821              \\
Thunderbird & —               & 19.532              & 4.069              \\
Windows     & 1.357           & 31.938              & 5.507              \\
Zookeeper   & 0.842           & 3.071               & 2.523              \\
\bottomrule
\end{tabular}
}
\vspace{-6pt}
\end{table}

%% file: tables/rq2_ablation_cr.tex
\begin{table}[htbp]
\centering
\small
\caption{Ablation study results}
\vspace{-6pt}
\label{tab:r2_cr}
\setlength{\tabcolsep}{4pt}
\renewcommand{\arraystretch}{1.15}
\begin{tabular}{@{}c|cc|cc|cc@{}}
\toprule
\multirow{2}{*}{\textbf{Dataset}} & \multicolumn{2}{c|}{\textbf{LogShrink}} & \multicolumn{2}{c|}{\textbf{w/o Sampler}} & \multicolumn{2}{c}{\textbf{w/o Analyzer}} \\ \cmidrule(l){2-7} 
            & CR      & CS    & CR      & CS    & CR      & CS     \\ \midrule
Android     & 21.857  & 5.347 & 21.784  & 4.354 & 20.804  & 7.048  \\
Apache      & 55.940  & 1.880 & 56.075  & 1.565 & 42.803  & 2.286  \\
BGL         & 42.385  & 2.519 & 43.489  & 1.767 & 32.965  & 5.283  \\
Hadoop      & 60.091  & 3.137 & 61.206  & 2.904 & 57.885  & 7.281  \\
HDFS        & 27.319  & 3.253 & 31.270  & 2.737 & 22.023  & 7.631  \\
HealthApp   & 39.072  & 2.064 & 39.726  & 1.569 & 28.015  & 4.276  \\
HPC         & 35.878  & 2.485 & 36.706  & 1.894 & 27.317  & 4.771  \\
Linux       & 29.252  & 1.307 & 29.365  & 1.151 & 25.310  & 1.253  \\
Mac         & 39.860  & 2.572 & 39.333  & 2.258 & 35.183  & 3.878  \\
OpenSSH     & 103.175 & 3.409 & 101.727 & 2.675 & 71.874  & 5.382  \\
OpenStack   & 22.157  & 2.945 & 22.648  & 2.136 & 20.573  & 5.484  \\
Proxifier   & 27.029  & 1.315 & 28.061  & 1.119 & 26.262  & 1.265  \\
Spark       & 59.739  & 2.821 & 59.234  & 2.227 & 49.147  & 5.651  \\
Thunderbird & 48.434  & 4.069 & 45.367  & 3.289 & 40.130  & 6.047  \\
Windows     & 456.301 & 5.507 & 501.502 & 4.316 & 390.574 & 11.445 \\
Zookeeper   & 116.981 & 2.523 & 120.003 & 2.002 & 70.993  & 3.066  \\ \bottomrule
\multicolumn{7}{l}{%
}\\
\end{tabular}
\vspace{-6pt}
\end{table}

%% file: sections/discussion.tex






\section{Threats to Validity}
\label{sec:discussion}
We have identified the following threats to validity:


\textbf{Subject systems:} We only conduct our empirical study on three representative datasets. 
Also, our experiments are performed on a limited number of log datasets from 16 subject systems, which cannot represent all software systems. 
In the future, we plan to collect more log data and evaluate our methods on more software systems.

\textbf{Implementation:} 
The runtime performance of a program can differ across different programming languages. We have implemented LogShrink using Python, which prioritizes code readability over execution speed. 
According to a comparison of the speed of programming languages~\cite{plspeed}, 
the performance of C/C++ is 10 $\times$ as fast as Python.
As a result, the compression speed 
of LogShrink is slower compared to compressors~\cite{gzip,7za,bzip2,wei2021feasibility}, which are written in faster programming languages like C++. This language bias can affect the comparison of compression speed. To address this, we plan to optimize the LogShrink implementation in other faster programming languages, such as C++, in the future.

\textbf{Tool comparison:} In our evaluation, we compared LogShrink with three general-purpose compressors (i.e., lzma~\cite{7za}, gzip~\cite{gzip}, bzip2~\cite{bzip2}) and two state-of-the-art log-specific compressors (i.e., LogZip~\cite{liu2019logzip}, LogReducer~\cite{wei2021feasibility}). To reduce the threat from tool comparison, we directly use the code provided by their papers~\cite{lgozipcode, logreducercode}. Also, we use the popular implementations of the general-purpose compressors~\cite{7za, gzip, bzip2}.


%% file: main.bib
@misc{gzip,
  title = {The gzip home page},
  howpublished = "\url{https://www.gzip.org}",
  note = "[Online]",
  year = {2023}
}

@misc{bzip2,
  title = {The bzip2 home page},
  howpublished = "\url{https://sourceware.org/bzip2/}",
  note = "[Online]",
  year = {2023}
}

@misc{hadoop,
  title = {Hadoop},
  howpublished = "\url{https://hadoop.apache.org/}",
  note = "[Online]",
  year = {2021}
}

@misc{loghubdatasets,
  title = {LogHub Datasets},
  howpublished = "\url{https://zenodo.org/record/3227177}",
  note = "[Online]",
  year = {2023}
}

@misc{logreducercode,
  title = {Open source code of LogReducer},
  howpublished = "\url{https://github.com/THUBear-wjy/LogReducer}",
  note = "[Online]",
  year = {2023}
}

@misc{lgozipcode,
  title = {Open source code of LogZip},
  howpublished = "\url{https://github.com/logpai/logzip}",
  note = "[Online]",
  year = {2023}
}

@misc{tarfile,
  title = {Python Library tarfile},
  howpublished = "\url{https://docs.python.org/3/library/tarfile.html}",
  note = "[Online]",
  year = {2023}
}

@misc{7za,
  title = {7za tool},
  howpublished = "\url{https://linux.die.net/man/1/7za}",
  note = "[Online]",
  year = {2023}
}

@misc{bgl,
  title = {BlueGene/L message types},
  howpublished = "\url{https://www.usenix.org/cfdr-data#hpc4}",
  note = "[Online]",
  year = {2019}
}

@inproceedings{SCLTvaarandi2003data,
  title={A data clustering algorithm for mining patterns from event logs},
  author={Vaarandi, Risto},
  booktitle={IPOM'03: Proc. of the 3rd IEEE Workshop on IP Operations \& Management},
  pages={119--126},
  year={2003},
  organization={IEEE}
}

@inproceedings{he2017drain,
  title={Drain: An online log parsing approach with fixed depth tree},
  author={He, Pinjia and Zhu, Jieming and Zheng, Zibin and Lyu, Michael R},
  booktitle={ICWS'17: 2017 IEEE International Conference on Web Services},
  pages={33--40},
  year={2017},
  organization={IEEE}
}

@inproceedings{du2016spell,
  title={Spell: Streaming parsing of system event logs},
  author={Du, Min and Li, Feifei},
  booktitle={ICDM'16: Proc. of the 16th International Conference on Data Mining},
  pages={859--864},
  year={2016},
  organization={IEEE}
}

@inproceedings{du2017deeplog,
  title={Deeplog: Anomaly detection and diagnosis from system logs through deep learning},
  author={Du, Min and Li, Feifei and Zheng, Guineng and Srikumar, Vivek},
  booktitle={SIGSAC'17: Proc. of the 2017 ACM SIGSAC Conference on Computer and Communications Security},
  pages={1285--1298},
  year={2017},
  organization={ACM}
}

@inproceedings{ijcai2019-658,
  title={LogAnomaly: Unsupervised Detection of Sequential and Quantitative Anomalies in Unstructured Logs.},
  author={Meng, Weibin and Liu, Ying and Zhu, Yichen and Zhang, Shenglin and Pei, Dan and Liu, Yuqing and Chen, Yihao and Zhang, Ruizhi and Tao, Shimin and Sun, Pei and others},
  booktitle={IJCAI'19: Proc. of the 28th International Joint Conference on Artificial Intelligence},
  pages={4739--4745},
  year={2019}
}

@article{he2018directed,
  title={A directed acyclic graph approach to online log parsing},
  author={He, Pinjia and Zhu, Jieming and Xu, Pengcheng and Zheng, Zibin and Lyu, Michael R},
  journal={arXiv preprint arXiv:1806.04356},
  year={2018}
}

@inproceedings{yang2018nanolog,
  author       = {Stephen Yang and
                  Seo Jin Park and
                  John K. Ousterhout},
  title        = {NanoLog: {A} Nanosecond Scale Logging System},
  booktitle    = {2018 {USENIX} Annual Technical Conference, {USENIX} {ATC} 2018, Boston, MA, USA, July 11-13, 2018},
  pages        = {335--350},
  publisher    = {{USENIX} Association},
  year         = {2018},
  bibsource    = {dblp computer science bibliography, https://dblp.org}
}

@article{yuan2012improving,
  title={Improving software diagnosability via log enhancement},
  author={Yuan, Ding and Zheng, Jing and Park, Soyeon and Zhou, Yuanyuan and Savage, Stefan},
  journal={ACM Transactions on Computer Systems (TOCS)},
  volume={30},
  number={1},
  pages={1--28},
  year={2012},
  publisher={ACM New York, NY, USA}
}

@inproceedings{he2018characterizing,
  title={Characterizing the natural language descriptions in software logging statements},
  author={He, Pinjia and Chen, Zhuangbin and He, Shilin and Lyu, Michael R},
  booktitle={Proceedings of the 33rd ACM/IEEE International Conference on Automated Software Engineering},
  pages={178--189},
    publisher    = {{ACM}},
  year={2018}
}

@article{liu2019variables,
  title={Which variables should i log?},
  author={Liu, Zhongxin and Xia, Xin and Lo, David and Xing, Zhenchang and Hassan, Ahmed E and Li, Shanping},
  journal={IEEE Transactions on Software Engineering},
  volume={47},
  number={9},
  pages={2012--2031},
  year={2019},
  publisher={IEEE}
}

@inproceedings{hamooni2016logmine,
  title={Logmine: Fast pattern recognition for log analytics},
  author={Hamooni, Hossein and Debnath, Biplob and Xu, Jianwu and Zhang, Hui and Jiang, Guofei and Mueen, Abdullah},
  booktitle={CIKM'16: Proc. of the 25th ACM International on Conference on Information and Knowledge Management},
  pages={1573--1582},
  year={2016},
  organization={ACM}
}

@inproceedings{he2016experience,
  title={Experience report: system log analysis for anomaly detection},
  author={He, Shilin and Zhu, Jieming and He, Pinjia and Lyu, Michael R},
  booktitle={ISSRE'16: Proc. of the 27th International Symposium on Software Reliability Engineering},
  pages={207--218},
  year={2016},
  organization={IEEE}
}

@inproceedings{he2018identifying,
  title={Identifying impactful service system problems via log analysis},
  author={He, Shilin and Lin, Qingwei and Lou, Jian-Guang and Zhang, Hongyu and Lyu, Michael R and Zhang, Dongmei},
  booktitle={Proceedings of the 2018 26th ACM Joint Meeting on European Software Engineering Conference and Symposium on the Foundations of Software Engineering},
  pages={60--70},
  year={2018}
}

@inproceedings{li2020swisslog,
  title={SwissLog: Robust and Unified Deep Learning Based Log Anomaly Detection for Diverse Faults},
  author={Li, Xiaoyun and Chen, Pengfei and Jing, Linxiao and He, Zilong and Yu, Guangba},
  booktitle={2020 IEEE 31st International Symposium on Software Reliability Engineering (ISSRE)},
  pages={92--103},
  year={2020},
  organization={IEEE}
}

@article{li2022swisslog,
  author       = {Xiaoyun Li and
                  Pengfei Chen and
                  Linxiao Jing and
                  Zilong He and
                  Guangba Yu},
  title        = {SwissLog: Robust Anomaly Detection and Localization for Interleaved
                  Unstructured Logs},
  journal      = {{IEEE} Transactions on Dependable and Secure Computing},
  volume       = {20},
  number       = {4},
  pages        = {2762--2780},
  year         = {2023},
  bibsource    = {dblp computer science bibliography, https://dblp.org}
}

@inproceedings{li2022going,
  title={Going through the Life Cycle of Faults in Clouds: Guidelines on Fault Handling},
  author={Li, Xiaoyun and Yu, Guangba and Chen, Pengfei and Chen, Hongyang and Chen, Zhekang},
  booktitle={2022 IEEE 33rd International Symposium on Software Reliability Engineering (ISSRE)},
  pages={121--132},
  year={2022},
  organization={IEEE}
}

@inproceedings{zhou2019latent,
  title={Latent error prediction and fault localization for microservice applications by learning from system trace logs},
  author={Zhou, Xiang and Peng, Xin and Xie, Tao and Sun, Jun and Ji, Chao and Liu, Dewei and Xiang, Qilin and He, Chuan},
  booktitle={Proceedings of the 2019 27th ACM Joint Meeting on European Software Engineering Conference and Symposium on the Foundations of Software Engineering},
  pages={683--694},
  year={2019}
}

@inproceedings{zhang2022deeptralog,
  title={DeepTraLog: Trace-log combined microservice anomaly detection through graph-based deep learning},
  author={Zhang, Chenxi and Peng, Xin and Sha, Chaofeng and Zhang, Ke and Fu, Zhenqing and Wu, Xiya and Lin, Qingwei and Zhang, Dongmei},
  booktitle={Proceedings of the 44th International Conference on Software Engineering},
  pages={623--634},
  year={2022}
}

@inproceedings{yuan2012characterizing,
  title={Characterizing logging practices in open-source software},
  author={Yuan, Ding and Park, Soyeon and Zhou, Yuanyuan},
  booktitle={2012 34th International Conference on Software Engineering (ICSE)},
  pages={102--112},
  year={2012},
  organization={IEEE}
}

@inproceedings{fu2014developers,
  title={Where do developers log? an empirical study on logging practices in industry},
  author={Fu, Qiang and Zhu, Jieming and Hu, Wenlu and Lou, Jian-Guang and Ding, Rui and Lin, Qingwei and Zhang, Dongmei and Xie, Tao},
  booktitle={Companion Proceedings of the 36th International Conference on Software Engineering},
  pages={24--33},
publisher    = {{ACM}},
  year={2014}
}

@article{foalem2023studying,
  title={Studying Logging Practice in Machine Learning-based Applications},
  author={Foalem, Patrick Loic and Khomh, Foutse and Li, Heng},
  journal={arXiv preprint arXiv:2301.04234},
  year={2023}
}

@misc{plspeed,
  title = {Speed comparison of programming languages},
  howpublished = "\url{https://github.com/niklas-heer/speed-comparison}",
  note = "[Online]",
  year = {2023}
}

@misc{loggingpricing,
  title = {Logging Storage Pricing},
  howpublished = "\url{https://cloud.google.com/stackdriver/pricing}",
  note = "[Online]",
  year = {2023}
}

@article{huffman1952method,
  title={A Method for the Construction of Minimum-Redundancy Codes},
  author={Huffman, David A},
  journal={Proceedings of the IRE},
  volume={40},
  number={9},
  pages={1098--1101},
  year={1952},
  publisher={IEEE}
}

@article{witten1987arithmetic,
  title={Arithmetic Coding for Data Compression},
  author={Witten, Ian H and Neal, Radford M and Cleary, John G},
  journal={Communications of the ACM},
  volume={30},
  number={6},
  pages={520--540},
  year={1987},
  publisher={ACM New York, NY, USA}
}

@article{cleary1984data,
  title={Data compression using adaptive coding and partial string matching},
  author={Cleary, John and Witten, Ian},
  journal={IEEE transactions on Communications},
  volume={32},
  number={4},
  pages={396--402},
  year={1984},
  publisher={IEEE}
}

@inproceedings{goyal2018deepzip,
  author       = {Mohit Goyal and
                  Kedar Tatwawadi and
                  Shubham Chandak and
                  Idoia Ochoa},
  title        = {DeepZip: Lossless Data Compression Using Recurrent Neural Networks},
  booktitle    = {Data Compression Conference, {DCC} 2019},
  pages        = {575},
  publisher    = {{IEEE}},
  year         = {2019},
  bibsource    = {dblp computer science bibliography, https://dblp.org}
}

@inproceedings{liu2019logzip,
  author       = {Jinyang Liu and
                  Jieming Zhu and
                  Shilin He and
                  Pinjia He and
                  Zibin Zheng and
                  Michael R. Lyu},
  title        = {Logzip: Extracting Hidden Structures via Iterative Clustering for
                  Log Compression},
  booktitle    = {34th {IEEE/ACM} International Conference on Automated Software Engineering,
                  {ASE} 2019},
  pages        = {863--873},
  publisher    = {{IEEE}},
  year         = {2019},
  bibsource    = {dblp computer science bibliography, https://dblp.org}
}

@inproceedings{wei2021feasibility,
  title={On the Feasibility of Parser-based Log Compression in $\{$Large-Scale$\}$ Cloud Systems},
  author={Wei, Junyu and Zhang, Guangyan and Wang, Yang and Liu, Zhiwei and Zhu, Zhanyang and Chen, Junchao and Sun, Tingtao and Zhou, Qi},
  booktitle={19th USENIX Conference on File and Storage Technologies (FAST 21)},
    publisher    = {{USENIX} Association},
  pages={249--262},
  year={2021}
}

@article{yao2021improving,
  title={Improving state-of-the-art compression techniques for log management tools},
  author={Yao, Kundi and Sayagh, Mohammed and Shang, Weiyi and Hassan, Ahmed E},
  journal={IEEE Transactions on Software Engineering},
  volume={48},
  number={8},
  pages={2748--2760},
  year={2021},
  publisher={IEEE}
}

@inproceedings{wei2023loggrep,
  title={LogGrep: Fast and Cheap Cloud Log Storage by Exploiting both Static and Runtime Patterns},
  author={Wei, Junyu and Zhang, Guangyan and Chen, Junchao and Wang, Yang and Zheng, Weimin and Sun, Tingtao and Wu, Jiesheng and Jiang, Jiangwei},
  booktitle={Proceedings of the 18th European Conference on Computer Systems},
  pages={452--468},
  year={2023}
}

@inproceedings{ding2021elise,
  title={ELISE: A Storage Efficient Logging System Powered by Redundancy Reduction and Representation Learning.},
  author={Ding, Hailun and Yan, Shenao and Zhai, Juan and Ma, Shiqing},
  booktitle={USENIX Security Symposium},
  pages={3023--3040},
  year={2021}
}

@inproceedings{lin2015cowic,
  title={Cowic: A column-wise independent compression for log stream analysis},
  author={Lin, Hao and Zhou, Jingyu and Yao, Bin and Guo, Minyi and Li, Jie},
  booktitle={2015 15th IEEE/ACM International Symposium on Cluster, Cloud and Grid Computing},
  pages={21--30},
  year={2015},
  organization={IEEE}
}

@inproceedings{christensen2013adaptive,
  author       = {Robert Christensen and
                  Feifei Li},
  title        = {Adaptive log compression for massive log data},
  booktitle    = {Proceedings of the {ACM} {SIGMOD} International Conference on Management
                  of Data, {SIGMOD} 2013},
  pages        = {1283--1284},
  publisher    = {{ACM}},
  year         = {2013},
  bibsource    = {dblp computer science bibliography, https://dblp.org}
}

@inproceedings{feng2016mlc,
  title={MLC: An efficient multi-level log compression method for cloud backup systems},
  author={Feng, Bo and Wu, Chentao and Li, Jie},
  booktitle={2016 IEEE Trustcom/BigDataSE/ISPA},
  pages={1358--1365},
  year={2016},
  organization={IEEE}
}

@inproceedings{rodrigues2021clp,
  title={CLP: Efficient and Scalable Search on Compressed Text Logs.},
  author={Rodrigues, Kirk and Luo, Yu and Yuan, Ding},
  booktitle={OSDI},
  pages={183--198},
  year={2021}
}

@inproceedings{andrews1998testing,
  title={Testing using log file analysis: tools, methods, and issues},
  author={Andrews, James H},
  booktitle={Proceedings 13th IEEE International Conference on Automated Software Engineering (Cat. No. 98EX239)},
  pages={157--166},
  year={1998},
  organization={IEEE}
}

@inproceedings{chen2018automated,
  title={An automated approach to estimating code coverage measures via execution logs},
  author={Chen, Boyuan and Song, Jian and Xu, Peng and Hu, Xing and Jiang, Zhen Ming},
  booktitle={Proceedings of the 33rd ACM/IEEE International Conference on Automated Software Engineering},
  pages={305--316},
  year={2018}
}

@inproceedings{yao2018log4perf,
  title={Log4perf: Suggesting logging locations for web-based systems' performance monitoring},
  author={Yao, Kundi and B. de P{\'a}dua, Guilherme and Shang, Weiyi and Sporea, Steve and Toma, Andrei and Sajedi, Sarah},
  booktitle={Proceedings of the 2018 ACM/SPEC International Conference on Performance Engineering},
  pages={127--138},
  year={2018}
}

@inproceedings{agrawal2018log,
  title={Log-based cloud monitoring system for OpenStack},
  author={Agrawal, Vaibhav and Kotia, Devanjal and Moshirian, Kamelia and Kim, Mihui},
  booktitle={2018 IEEE Fourth International Conference on Big Data Computing Service and Applications (BigDataService)},
  pages={276--281},
  year={2018},
  organization={IEEE}
}

@inproceedings{zhang2020anomaly,
  title={Anomaly detection via mining numerical workflow relations from logs},
  author={Zhang, Bo and Zhang, Hongyu and Moscato, Pablo and Zhang, Aozhong},
  booktitle={2020 International Symposium on Reliable Distributed Systems (SRDS)},
  pages={195--204},
  year={2020},
  organization={IEEE}
}

@article{ziv1977universal,
  title={A Universal Algorithm for Sequential Data Compression},
  author={Ziv, Jacob and Lempel, Abraham},
  journal={IEEE Transactions on information theory},
  volume={23},
  number={3},
  pages={337--343},
  year={1977},
  publisher={IEEE}
}

@inproceedings{yu2023logreducer,
  title={LogReducer: Identify and Reduce Log Hotspots in Kernel on the Fly},
  author={Yu, Guangba and Chen, Pengfei and Li, Pairui and Weng, Tianjun and Zheng, Haibing and Deng, Yuetang and Zheng, Zibin},
  booktitle={2023 IEEE/ACM 45th International Conference on Software Engineering (ICSE)},
  pages={1763--1775},
  year={2023},
  organization={IEEE}
}

@techreport{deutsch1996gzip,
  title={GZIP file format specification version 4.3},
  author={Deutsch, Peter},
  year={1996}
}

@misc{elk-stack,
  title = {ELK Stack},
  howpublished = "\url{https://www.elastic.co/elastic-stack}",
  note = "[Online]",
  year = {2023}
}

@article{yao2020study,
  title={A study of the performance of general compressors on log files},
  author={Yao, Kundi and Li, Heng and Shang, Weiyi and Hassan, Ahmed E},
  journal={Empirical Software Engineering},
  volume={25},
  pages={3043--3085},
  year={2020},
  publisher={Springer}
}

@article{sanger2021scope,
  title={Scope of Russian hacking becomes clear: multiple US agencies were hit},
  author={Sanger, David E and Perlroth, Nicole and Schmitt, Eric},
  journal={The New York Times},
  year={2021}
}

@article{bing2020suspected,
  title={Suspected Russian hackers spied on US Treasury emails-sources},
  author={Bing, Christopher},
  journal={Reuters, Dec},
  volume={13},
  year={2020}
}

@inproceedings{ding2015log2,
  author       = {Rui Ding and
                  Hucheng Zhou and
                  Jian{-}Guang Lou and
                  Hongyu Zhang and
                  Qingwei Lin and
                  Qiang Fu and
                  Dongmei Zhang and
                  Tao Xie},
  title        = {Log2: {A} Cost-Aware Logging Mechanism for Performance Diagnosis},
  booktitle    = {2015 {USENIX} Annual Technical Conference, {USENIX} {ATC} '15},
  pages        = {139--150},
  publisher    = {{USENIX} Association},
  year         = {2015},
  bibsource    = {dblp computer science bibliography, https://dblp.org}
}

@inproceedings{zhao2017log20,
  title={Log20: Fully automated optimal placement of log printing statements under specified overhead threshold},
  author={Zhao, Xu and Rodrigues, Kirk and Luo, Yu and Stumm, Michael and Yuan, Ding and Zhou, Yuanyuan},
  booktitle={Proceedings of the 26th Symposium on Operating Systems Principles},
  pages={565--581},
  year={2017}
}

@inproceedings{adiego2004lempel,
  title={Lempel-Ziv compression of structured text},
  author={Adiego, Joaqu{\'\i}n and Navarro, Gonzalo and de la Fuente, Pablo},
  booktitle={Data Compression Conference, 2004. Proceedings. DCC 2004},
  pages={112--121},
  year={2004},
  organization={IEEE}
}

@inproceedings{nevill1996compressing,
  title={Compressing semi-structured text using hierarchical phrase identifications},
  author={Nevill-Manning, Craig G and Witten, Ian H and Olsen, Dan R},
  booktitle={Proceedings of Data Compression Conference-DCC'96},
  pages={63--72},
  year={1996},
  organization={IEEE}
}

@article{zhang2023semi,
  title={Semi-supervised and unsupervised anomaly detection by mining numerical workflow relations from system logs},
  author={Zhang, Bo and Zhang, Hongyu and Le, Van-Hoang and Moscato, Pablo and Zhang, Aozhong},
  journal={Automated Software Engineering},
  volume={30},
  number={1},
  pages={4},
  year={2023},
  publisher={Springer}
}

@inproceedings{le2022log,
  title={Log-based anomaly detection with deep learning: How far are we?},
  author={Le, Van-Hoang and Zhang, Hongyu},
  booktitle={Proceedings of the 44th international conference on software engineering},
  pages={1356--1367},
  year={2022}
}

@inproceedings{DBLP:conf/icse/LeZ23,
  author       = {Van{-}Hoang Le and
                  Hongyu Zhang},
  title        = {Log Parsing with Prompt-based Few-shot Learning},
  booktitle    = {45th {IEEE/ACM} International Conference on Software Engineering},
  pages        = {2438--2449},
  publisher    = {{IEEE}},
  year         = {2023},
  bibsource    = {dblp computer science bibliography, https://dblp.org}
}
